\documentclass{aa}  

\usepackage{graphicx}
\usepackage{txfonts}
\usepackage[final=true, pageanchor=true, colorlinks=true, breaklinks=true, linkcolor=blue, citecolor=blue, urlcolor=blue, pdfpagemode=UseNone]{hyperref}
\begin{document}

\title{Evolution of the flow field in decaying active regions}

\subtitle{Transition from a moat flow to a supergranular flow}

 \author{H. Strecker \and
 	N. Bello Gonz\'alez} 
	
\institute{Kiepenheuer-Institut f{\"u}r Sonnenphysik,
	Sch{\"o}neckstr.\ 6,
   	79104 Freiburg,
   	Germany\\
	\email{[strecker;nbello]@leibniz-kis.de}
	}
 
\date{Received  October 24, 2017; accepted September, 21 2018}
\abstract
%Context
{Fully fledged sunspots are known to be surrounded by a radial outflow called the moat flow.}
%Aims:
{We investigate the evolution of the horizontal flow field around sunspots during their decay, that is, from fully fledged to naked spots, after they loose the penumbra, to the remnant region after the spot has fully dissolved.}
%Methods:
{We analysed the extension and horizontal velocity of the flow field around eight sunspots using SDO/HMI Doppler maps. By assuming a radially symmetrical flow field, the applied analysis method determines the radial dependence of the azimuthally averaged flow field. For comparison, we studied the flow in supergranules using the same technique.}
%Results:
{All investigated, fully fledged sunspots are surrounded by a flow field whose horizontal velocity profile decreases continuously from 881\,m\,s$^{-1}$ at 1.1\,Mm off the spot boundary, down to 199\,m\,s$^{-1}$ at a mean distance of 11.9\,Mm to that boundary, in agreement with other studies. 
Once the penumbra is fully dissolved, however, the velocity profile of the flow changes: The horizontal velocity increases with increasing distance to the spot boundary until a maximum value of about 398\,m\,s$^{-1}$ is reached. Then, the horizontal velocity decreases for farther distances to the spot boundary. 
In supergranules, the horizontal velocity increases with increasing distance to their centre up to a mean maximum velocity of 355\,m\,s$^{-1}$. For larger distances, the horizontal velocity decreases. We thus find that the velocity profile of naked sunspots resembles that of supergranular flows.
The evolution of the flow field around individual sunspots is influenced by the way the sunspot decays and by the interplay with the surrounding flow areas.}
%Conclusions:
{Observations of the flow around eight decaying sunspots suggest that as long as penumbrae are present, sunspots with their moat cell are embedded in network cells. The disappearance of the penumbra (and consequently the moat flow) and the competing surrounding supergranular cells, both have a significant role in the evolution of the flow field: The moat cell transforms into a supergranule, which hosts the remaining naked spot.}
\maketitle
%
%________________________________________________________________
%
\section{Introduction}\label{introduction}
%===================================================================================
%                                CHARACTERISTIC of MOAT FLOW
A fully evolved sunspot with a penumbra is usually surrounded by an annular cell, the so-called moat cell. This region shows a horizontal-radial flow of gas away from the spot. \citet{sheeley1972} first analysed the flow around sunspots and described its annular form. The extension of the flow, as measured from the outer spot boundary, is stated to be in a range of 10--22\,Mm (\citealt{sheeley1972,muller1987,shine1987,brickhouse1988,balthasar2010,verma2018}). \citet{loehner-boettcher2013} studied the flow field of 31 H-class sunspots in HMI Doppler maps and obtained extension values in the range of 5--15\,Mm with a mean value of 9.2\,Mm. Time-distant helioseismic inversions by \citet{svanda2014} lead to moat flow extensions of 12\,Mm for an average over 104\,H-class spots. \citet{rempel2015} simulated two sunspots, one with a fully developed penumbra and one naked sunspot that had just lost its penumbra. He found an extension of 10\,Mm of the flow field around both sunspots.

\citet{sheeley1972} measured horizontal flow velocities of 0.5--1\,km\,s\textsuperscript{-1}. The comparison of several analysis methods (e.g. tracking granules and Doppler shifts) by \citealt{brickhouse1988} led to flow velocities in the range of 0.2--1.4\,km\,s$^{-1}$. \citet{muller1987} described the velocity of the outflow motion as a function of the distance to the sunspot, with the velocity slowing down near the outer edge of the flow region. Using local correlation tracking, \citet{balthasar2013} obtained a decreasing velocity from the spot boundary with 1\,km\,s$^{-1}$ down to 0.2\,km\,s$^{-1}$ at the outer edge of the moat flow region. \citet{loehner-boettcher2013} measured velocities of 0.8--1.2\,km\,s$^{-1}$ just outside the sunspot with a monotonic decrease to a mean value of 0.18\,km\,s$^{-1}$ at the outer edge of the moat cell where the noise level is reached. The moat flow around the fully developed sunspot simulated by \citet{rempel2015} shows a similar velocity profile. The horizontal velocity drops from 1.5\,km\,s$^{-1}$ at the boundary of the sunspot down to 0.2\,km\,s$^{-1}$ at a distance of $R\,=\,R_{\mathrm{spot}}$\,+\,10\,Mm from the spot centre. The flow field enclosing the simulated decaying sunspot displays a different profile for the horizontal velocity. At a distance of 1\,Mm to the spot boundary, the naked spot is surrounded by an outflow whose horizontal velocity increases up to 0.5\,km\,s$^{-1}$ to later decrease until it drops below 0.2\,km\,s$^{-1}$ at a distance of 10\,Mm from the outer edge of the spot.

%===================================================================================
%                                ORIGIN of MOAT FLOW
The mechanism driving the moat flow is a controversial topic in solar physics. \citet{vargasdominguez2008} described the unmagnetised moat flow as an extension of the magnetised Evershed flow. This flow is observed as a radially outward directed horizontal flow in the photospheric layers of the penumbra (\citealt{evershed1909}) of magnetised plasma. The main argument for a connection of both flows is the non-existence of the moat flow when a sunspot does not have a penumbra. \citet{vargasdominguez2007} analysed spots with different penumbral configurations and found that the moat only appears when a penumbra is present. The flow can only be measured in the direction of the penumbra filaments and is not detectable in directions transverse to them. 

The idea of a connection between the Evershed flow and the moat flow is rejected by the observation of an outflow after the penumbra has dissolved and because the Evershed flow is magnetised, whereas the moat flow is a mainly unmagnetised flow (\citealt{deng2007,verma2012,balthasar2013}). \citet{pardon1979} found a persisting flow in the remnant region two days after the spots had disappeared. They described the sunspot in its final stage to be surrounded by a stable cell that became unstable and collapsed when the spot was gone. According to \citet{deng2007}, the symmetry of the flow pattern becomes lost when the sunspot decays. They observed a decrease in motion when the penumbra had dissolved. In addition, the outward motion is not in the immediate surroundings of the naked spot, but is separated by an inward motion similar to that reported by \citet{rempel2015} in his simulated naked spot. \cite{balthasar2013} measured higher velocities of the outflow on the penumbral side of a decaying sunspot compared to the flow velocity on the side where the penumbra had already dissolved.

\citet{meyer1974} described the moat flow as a horizontal outflow resulting from the rising of gas beneath the penumbra that bends horizontally as a result of the shape of the penumbra. \citet{nye1988} developed a model based on this idea: The flow of energy to the surface is blocked, causing a heating of gas beneath the sunspot. The rising gas cannot be efficiently transported because the strong magnetic field of the sunspot prevents this, which leads to an increasing gas pressure below the spot. Consequently, the gas rises, but is blocked by the funnel shape of the penumbra and develops a strong horizontal component. In this model, the velocity of the flow depends on the depth of the penumbra. \citet{sobotka2007} found the moat size to be mostly independent of the spot radius. Instead, it seems to depend on the evolutionary stage of the sunspot. The investigation of \citet{loehner-boettcher2013} also showed no dependence between the spot radius and the extension of the moat flow and supported the model by \citet{nye1988}. \citet{rempel2015} found clear evidence in his simulations for the decoupled origin of the moat and Evershed flows. The outflow around the sunspots is caused by the perturbation of the up- and downflow balance that is due to the presence of the sunspot: Because around spots, cooler downflows are reduced, the average temperature rises, causing a vertical upflow close to the spot that becomes visible as the moat flow when it reaches the photosphere.

%===================================================================================
%                                SUNSPOT DECAY
The lifetimes of sunspots can vary from several days to several weeks. In a group of sunspots, the follower spot seldom achieves a stable configuration and is normally destroyed within days, while the leader spot can evolve into a round and stable (H-class) sunspot whose decaying process can last up to several weeks (\citealt{bumba1963}). The slow decay of a sunspot is a diffusive process, which means a gradually shrinking by erosion at its boundary (\citealt{stix2004}). Nevertheless, \citet{sheeley2017} observed decaying sunspots that lost their round shape and became more elongated. Then, part of the flux separated from the spot, and the spot is reshaped into a smaller roundish spot. The authors concluded that sunspot decay occurs in a series of episodic bursts. \citet{bellotrubio2008} observed that the loss of the penumbra takes three days, in which the umbra splits and only three distinct finger-like structures remain, which extend from the umbra into the quiet Sun. \citet{deng2007} described the decay of the penumbra in relation with a change in and around flux concentrations of the local flow field. The photospheric plasma flow takes action and advects the flux, which contributes to the decay process of the sunspot. The spreading of magnetic flux over a larger area leads to the disappearance of the dark area in intensity maps, but does not remove magnetic flux from the photosphere (\citealt{martinezpillet2002}). Convective flows shape the field and destroy the annular pattern around the spot. Surrounding supergranular flows are able to carve out field-free regions and sweep away the flux to the cell edges. This leads to a fragmentation of the magnetic field, and in connection with magnetic flux of opposite polarity, it leads to flux cancellation (\citealt{vandriel-gesztelyi2015}).

%===================================================================================
%                                SUPERGRANULES
\citet{simon1964} described supergranulation as a system of atmospheric currents in the photosphere. The currents form a cellular network, which is visible in Doppler maps after the reduction of other larger scale flows (e.g. differential rotation and convective blueshift). Inside the cells, the flow is radially directed from the cell centre to its boundary. The diameters of supergranules are in the range of 10\,Mm up to 45\,Mm (\citealt{simon1964,roudier2014,orozcosuarez2012}). Depending on the method, different values are measured (\citealt{hirzberger2008}). A mean horizontal velocity of 0.4\,km\,s$^{-1}$ in supergranules was measured by \citet{simon1964}. \citet{orozcosuarez2012} investigated supergranular convective flows using Fourier local correlation tracking and intergranular magnetic elements. The flow velocity in supergranules increases outwards for larger distances from the centre. After reaching a maximum of 0.35\,km\,s$^{-1}$, the velocity decreases monotonically. By averaging over 222\,976 supergranular cells and applying time-distant helioseismic inversions, \citet{svanda2014} found a symmetrical flow in supergranules, which is directed radially away from the centre of the cells to its periphery, with horizontal velocities in the range of 0.3 to 0.6\,km\,s$^{-1}$. Their velocity profile is found to increase with increasing distances to the supergranule centre up to a maximum value from where on a continuous decrease for larger distances starts. According to \cite{simon1968}, the non-stationary cells only survive their turnover time, then the flow lapses into disorder. Supergranular cells have a mean lifetime of 1.5\,days, which can extend up to 4\,days (\citealt{roudier2014,hirzberger2008}). \citet{derosa2004} observed the creation of supergranules due to fragmentation or merging of older cells and stated that each supergranular cell takes part in a minimum of one merging or splitting event during its lifetime. This interaction seems to be the preferred mode of evolution of the observed supergranular pattern. The mechanism of advecting magnetic field elements to the network and the similar appearance suggests a relation between the supergranular pattern and the quiet-Sun magnetic network, as was recognised by \citet{simon1964}. \citet{derosa2004} observed a strengthening or weakening of the network lanes due to splitting or merging of supergranules. Although there is no definite evidence of a one-to-one relation (see review by \citet{rieutord2010} and references therein), in the large-scale picture, the supergranular pattern can be equated with the magnetic network. 
%===================================================================================
%                                RELATION MF and SUPERGRANULES
The horizontal, radially outward directed gas motion of the moat flow, which is visible in Doppler maps, resembles the characteristics of supergranular flows. In addition, magnetic lanes can form around the moat cell due to magnetic features (MMFs), which cross the moat and conglomerate, resembling the magnetic network. This relation has been described by \cite{simon1964}, \citet{sheeley1972}, and \citet{vrabec1974}, for example, who proposed the sunspot to be sitting in the centre of a supergranule. \citet{meyer1974} concluded from the observations that sunspots are related to supergranular convection, but these two flows should be distinguished because of their difference in size and occurrence by taking into account the unique relation of the moat flow to its sunspot. \cite{svanda2014} found the moat flow to be asymmetric, while flows in supergranules are symmetric, but the investigated cells approximately show the same size. In addition, the authors described the moat cell as a downflow region, while the motion in supergranules shows an upflow$-$downflow behaviour. Various investigations have been carried out to study the differences and similarities between the moat flow and supergranules (see e.g. \cite{sobotka2007} and the reviews by \cite{solanki2003} and \citet{rieutord2010} and references therein). An overview of the characteristics of the moat flow and supergranules is given in Table\,\ref{tab:mf_supergranules}. 

%%%%%%%%%%%  TABLE 1 %%%%%%%%%%%%%%%
\begin{table}[!ht]
        \centering
        \begin{small}
                \caption{Characteristics of the moat flow and supergranules summarised as an overview.}
                \begin{tabular}{c c c}\hline
                        &Supergranule&Moat flow\\ \hline\hline
                        Extension&5--22.5\,Mm\textsuperscript{(5,8)}&5--20\,Mm (including spot)\textsuperscript{(2,4,7,10)}\\
                        (radius)&mean: 12.5\,Mm&mean: 9\,Mm (radius w/o spot)\\\hline
                        Velocity&0.3--0.6\,km\,s\textsuperscript{-1} \textsuperscript{(5,9,10)}&1.4--0.18\,km\,s\textsuperscript{-1}\textsuperscript{(1,4,7)}\\
                        &mean: 0.4\,km\,s\textsuperscript{-1}&mean: 0.55\,km\,s\textsuperscript{-1}\\\hline
                        Lifetime&mean: 1.5\,days\textsuperscript{(3,8)}&several days\textsuperscript{(6)}\\
                        &max: 4\,days&as long as an H-class sunspot\\\hline
                \end{tabular}
        \end{small}
        \tablebib{(1)~\citet{balthasar2013}; (2)~\citet{brickhouse1988}; (3)~\citet{hirzberger2008}; (4)~\citet{loehner-boettcher2013}; (5)~\cite{orozcosuarez2012}; (6)~\citet{pardon1979}; (7)~\citet{rempel2015}; (8)~\citet{roudier2014}; (9)~\citet{simon1964}; (10)~\citet{svanda2014}} \label{tab:mf_supergranules}
\end{table}
%%%%%%%%%%%  TABLE 1 %%%%%%%%%%%%%%%
In our analysis, we study the characteristics of the flow field in which a sunspot is embedded for different stages of the sunspot evolution. We start the analysis for fully fledged H-class sunspots and follow their evolution until they dissolve. The paper is organised as follows: In Sect.\,\ref{sec:select_ana} we describe the selection criteria and analyse the data. We apply the method not only to the flow region around sunspots (Sect.\,\ref{sec:mf}), but also to study the flow in supergranules (Sect.\,\ref{sec:sgs}). In Sect.\,\ref{sec:disc_conc} we discuss the results, and we conclude in Sect.\,\ref{sec:concl}.
%===================================================================================
 %                              OBSERVATIONS and DATA ANALYSIS
%===================================================================================
\section{Data selection and data analysis}\label{sec:select_ana}
We analyse the evolution of the flow field around fully fledged sunspots throughout their decay as well as the flow field in supergranules. For this purpose, we used Doppler maps from the Solar Dynamics Observatory (SDO) (\citealt{pesnell2012}). SDO data allow for continuous observations over long time ranges, which is necessary to follow the evolution of a sunspot and its surrounding area. We selected eight sunspots according to two criteria:
\begin{enumerate}
\item The sunspots should be fully fledged and roundish (H-class sunspots) when they appear on the eastern limb of the solar disc.
\item The sunspots should decay before reaching the western limb.
\end{enumerate}
We used the sub-frame data provided by the HMI/AIA Joint Science Operations Center (JSOC)\footnote{\label{ftn:jsoc}\href{http://jsoc.stanford.edu/}{http://jsoc.stanford.edu/}}. The provided patches are generated by taking into account the NOAA number. In addition, the sub-frames are already rotated to switch the north-south orientation of the original data\footnote{\label{ftn:rotsat}Full disc data of the Helioseismic and Magnetic Imager (HMI) (\citet{schou2012}) show a rotation angle of 180$^{\circ}$.} . Full-disc data of the Helioseismic and Magnetic Imager (HMI) (\citet{schou2012}) show a rotation angle of 180$^{\circ}$. To ensure that we have the whole flow field surrounding the spot in the field of view (FOV), we chose a data size of $500\times500$\,pix$^2$, that is, $252\times 252$\,arcsec$^2$. 

The analysis method of our investigation is based on the work by \cite{loehner-boettcher2013}. They analysed the moat flow around 31  stable sunspots at a given heliocentric angle. We present here a brief summary of the analysis and refer to \cite{loehner-boettcher2013}, \cite{strecker2015}, and \cite{strecker2016} for a more detailed description of the method. In the latter, the method by \cite{loehner-boettcher2013} is extended to include the variation of the heliocentric angle during the passage of sunspots across the solar disc. 

%===================================================================================
%                                Reduction of Doppler maps
\subsection{Reduction of Doppler maps}
We used the HMI 720\,s Doppler maps provided by JSOC\footnote{\label{ftn:jsocd}\href{http://jsoc.stanford.edu/ajax/exportdata.html}{http://jsoc.stanford.edu/ajax/exportdata.html}}. These are low-noise data generated by combining filtergrams recorded over a time interval of 1260\,s. Several velocity patterns, which are superimposed in one Doppler map, have to be taken into account and were removed in order to properly study the flow field around sunspots. These velocity components are 
\begin{enumerate}
        \item the radial and tangential velocity components due to the orbital motion of the satellite, as given in the fits headers of the HMI files;
        \item the differential rotation of the Sun for the (time-dependent) B$_0$ inclination angle;
        \item the centre-to-limb variation of the convective blueshift; and
        \item a residual pattern caused by instrumental effects (generated with data from 2013 and 2014).
\end{enumerate} 

To generate the models of these four velocity components, we followed \cite{loehner-boettcher2013}. We obtained a different residual pattern (\citealt{strecker2015}), as our models are generated with data from 2013 and 2014. For a further reduction of the noise level and the effect of p-modes, we followed \cite{loehner-boettcher2013} and created 3\,h time averages (out of 15 successive Doppler maps).
%===================================================================================
%                                Localisation of analysis region
\subsection{Localisation of the region of interest}
For consistency, the simultaneous intensity maps and LOS-magnetograms were also averaged over three\,hours. The fomer were used to localise the sunspot as the centre of gravity and to determine the outer boundary of the sunspot by an intensity threshold of the spot compared to the surrounding region. 
After the disappearance of the spot in the intensity maps due to sunspot decay, an approximate localisation of the centre of the flow field by eye by taking into account the remaining flux in the LOS magnetograms. The obtained coordinates were used to crop out a field of view of $200\times200$\,pix$^2$ in the Doppler maps. The centre of the flow-field measurements is thus defined as the centre of the spot (as long as it is present) and located in the centre of the image. 
Animations available as online material show the evolution of each spot as seen in intensity, magnetograms, and Doppler maps (see Appendix\,\ref{app:spots} and Figs.\,\ref{fig:ar1_fig} - \ref{fig:ar8_fig} therein for temporal snapshots for each spot). The spot boundary is outlined by a red contour at each time. The position of the maximum horizontal velocity and extension of the horizontal flow, that is, the cell size as determined with our analysis method, are marked by green and pink contours, respectively. These allow following the spot evolution from fully fledged to vanished. To study supergranules, we defined their centre by hand as the centre of the magnetic network cells, as seen in the LOS magnetograms (see Fig.\,\ref{fig:sg_fig} in Appendix\,\ref{app:sg}) .

%===================================================================================
%                                Analysis of the flow field
\subsection{Analysis of the flow field}\label{sec:anaflow}
Owing to the spherical surface of the Sun, the velocities of horizontal flows on the solar surface such as the Evershed flow or the moat flow are redshifted for flows drifting away from disc centre and blueshifted for flows directed towards disc centre. We follow \cite{loehner-boettcher2013} in their analysis and assumed circular sunspots surrounded by an axially symmetrical flow field, which should also hold for supergranules. This allowed us to determine azimuthally averaged flow properties (\citealt{schlichenmaier2000}) by reading out the LOS velocity along circles with different radii around the centre of the sunspot or supergranule. Sunspots and so the surrounding flow field seem to have an elliptical shape off disc centre because of foreshortening effects. Therefore, the circles to read out the LOS velocity become elliptical according to the heliocentric angle, which defines the position of the region of interest on the solar disc. Based on the work by \cite{loehner-boettcher2013} (their Eq.\,5), the LOS horizontal velocity component $\upsilon^{\mathrm{LOS}}_\mathrm{h}$ for radial pixel step is obtained as the amplitude of the LOS velocity along the ellipse. For the LOS vertical component, which can be obtained as the offset of the fit, an accurate value cannot be obtained as it is strongly influenced by the residual velocities in the HMI Doppler maps. To compare the horizontal flow velocities for different heliocentric angles ($\theta$), $\upsilon^{\mathrm{LOS}}_\mathrm{h}(R,\theta)$, with $R$ the radial distance, they have to be independent of LOS projection effects. This is reached by 
\begin{equation}\upsilon_\mathrm{h}(R)=\frac{{\upsilon^{\mathrm{LOS}}_\mathrm{h}(R, \theta)}}{\sin(\theta)}\label{vh}.\end{equation}

Following \cite{loehner-boettcher2013}, we defined the size of the flow region at the point where the root mean square (rms) exceeds the LOS horizontal velocity component $\upsilon^{\mathrm{LOS}}_\mathrm{h}\leq\sigma_{\mathrm{rms}}$. The rms gives the azimuthal velocity fluctuation around the sine fit $f(\phi_i)$ and was computed for all distances $R\,=\,R_{\mathrm{spot}}\,+\,r_\mathrm{F}$ with $r_{\mathrm{F}}\ge3\,$pixel from the spot boundary:
\begin{equation}\sigma_{\mathrm{rms}}(R)=\sqrt{\frac{1}{n}\cdot\sum_{i=1}^{n}(\upsilon^{LOS}(R,\phi_i)-f(\phi_i))^2}.\end{equation}

\begin{figure*}[!h]
\includegraphics[width=1.\textwidth]{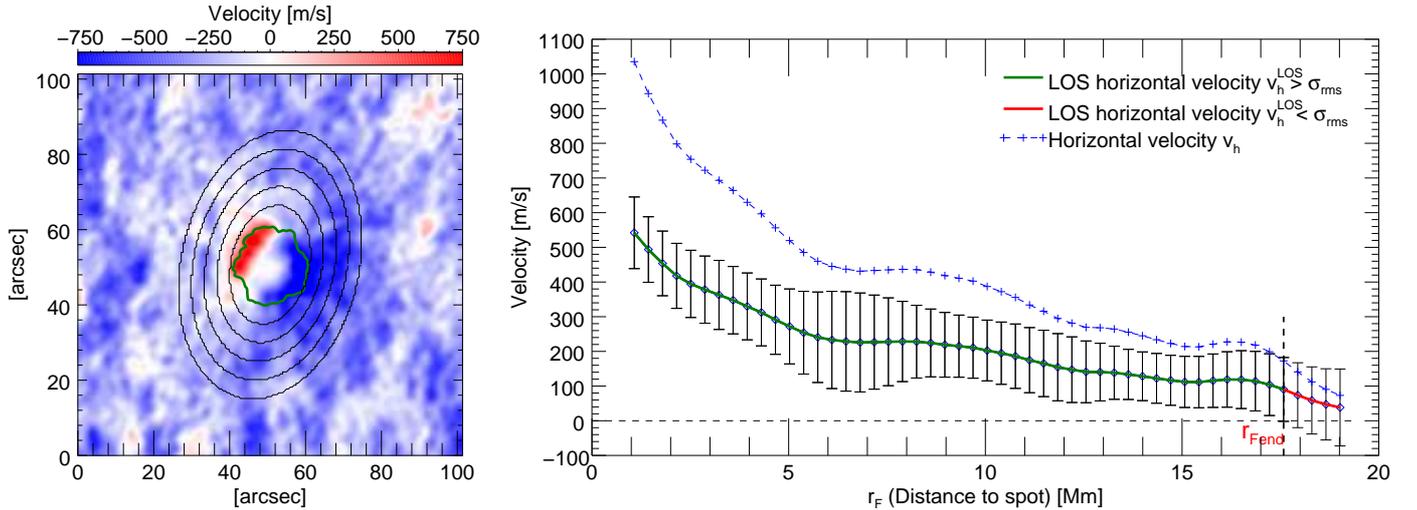}
\caption{Doppler map (left panel) of AR11641 on 2 January 2013 at 07:30\,UT. The green line displays the outline of the spot, and exemplary ellipses, along which the velocity values are read out, are plotted in black. The extension of the flow field is delimited by velocities higher than the $rms$ values (green curve in the right panel). The $rms$ values are plotted as error bars. The blue curve represents the velocity profile corrected for projection effects.}
\label{fig:velocityprofile}
\end{figure*}

Figure\,\ref{fig:velocityprofile} displays a snapshot of AR11641 (2 January 2013 at 07:30\,UT) as an example of the output of the analysis method:
In the left panel, a Doppler map is overplotted, with the green contour outlining the spot boundary as seen in intensity, and the various ellipses for different radii ($r_\mathrm{F}$) along which the flow field around the spot is measured. The right panel displays the variation of the LOS horizontal velocity with the radial distance to the spot boundary. The green curve displays LOS horizontal velocities higher than the rms value (error bars). The red curve displays LOS horizontal velocities that can no longer be detected above the noise (rms) level. This transition (from {\em green} to {\em red}) defines the outer boundary of the flow field (\citealt{loehner-boettcher2013}). The blue curve displays the horizontal velocity corrected for projection effects (Eq.\,1). The information on the maximum horizontal velocity and the extension of the flow were then extracted from $\upsilon_\mathrm{h}$ (independent of heliocentric angle) and $\upsilon^{\mathrm{LOS}}_\mathrm{h}$, respectively, for all studied sunspots and at different stages during their evolution across the disc.

%%%%%%%%%%%  FIGURE 1 %%%%%%%%%%%%%%%
%%%%%%%%%%%  FIGURE 1 %%%%%%%%%%%%%%%
The sunspots we studied do not keep their roundish shape during their decaying process. We defined the radius of the flow field to be $r_\mathrm{F}=0$\,pix at the spot boundary in intensity maps. To take into account the varying shape of the spot during its decaying process, which can lead to an underestimation of the radius by the automatic procedure, we started our analysis at a distance of $r_\mathrm{F}=3\,\mathrm{pix}\approx1.1$\,Mm from the sunspot boundary and took 51\,pixel steps to measure the velocity across the whole flow field. As the spot morphology changes, it also influences the morphology of the surrounding flow field. We took this aspect into account by examining the sunspot's magnetic structure and its surroundings in the corresponding LOS magnetograms, as we describe in the next section.
This is also illustrated in the animations provided in Appendix\,\ref{app:spots}.

%===================================================================================
%                                RESULTS
%===================================================================================
\section{Results}\label{sec:results}
In the first part of this section, we describe the results of the analysis of the flow field around decaying sunspots. The second part contains the results of the analysis of the flow in supergranules.
%===================================================================================
%                                Flow field around sunspots
\subsection{Flow field around sunspots}\label{sec:mf}
We analysed eight sunspots in different time spans, covering their evolution from fully fledged to naked sunspots to their eventual disappearance on the disc, when observable. We generated 3\,h data averages, which provided us with a set of eight data points per day. As required, all sunspots had fully fledged penumbrae at the beginning of the analysis. Table\,\ref{tab:list_sunspots} lists the main results for the analysed active regions. Columns.\,3 to 5 contain maximum horizontal velocity values for the three decaying stages of each spot. Columns\,6 to 8 contain the extension of the flow field also for the three decaying stages of each spot. The values in Cols.\,3 and 6 display averages over the first day in the analysis, that is, over eight samples in the fully fledged sunspot phase. Columns\,4 and 7 are time averages over the naked-spot phase, while Cols.\,5 and 8 are averages over the remaining analysis period, that is, after sunspot disappearance. We note that values in Cols.\,4 and 7 and 5 and 8 have different averaging periods depending on the evolution of each spot.
%%%%%%%%%%%  TABLE 2 %%%%%%%%%%%%%%%
\begin{table*}[!h]
        \centering 
        \caption{NOAA number of the investigated active regions with the extension of the flow field from the sunspot boundary $\bar r_{\mathrm{F_{end}}}$ and selected velocity values.}
        \resizebox{\linewidth}{!}{\begin{tabular}{c c c c c c c c} \hline
        AR      &       Period  &       $\bar \upsilon_\mathrm{h}$(3\,pixel) [m\,s\textsuperscript{-1}]        &       $\bar \upsilon^{\mathrm{max}}_\mathrm{h}(r_{\mathrm{F}})$ [m\,s\textsuperscript{-1}] &       $\bar \upsilon^{\mathrm{max}}_\mathrm{h}(r_{\mathrm{F}})$[m\,s\textsuperscript{-1}]  &       $\bar r_{\mathrm{F_{end}}}$ [Mm]      &       $\bar r_{\mathrm{F_{end}}}$ [Mm]        &       $\bar r_{\mathrm{F_{end}}}$ [Mm]      \\
                        &               &       (1\textsuperscript{st}\,day)    &       (w/o penumbra)       &       (w/o spot)      &       (1\textsuperscript{st}\,day)    &       (w/o penumbra)       &       (w/o spot)      \\\hline\hline
                11676   &       19.--27.02.2013 &       677     &       373     &       237     &       14.5    &       10.43   &       6.6     \\
                11641   &       02.--08.01.2013 &       838     &       488     &       342     &       14.8    &       12.9    &       13.1    \\
                11646   &       05.--10.01.2013 &       957     &       400     &       221     &       9.4     &       8.0     &       4.3     \\
                11841   &       13.--21.09.2013         &       711     &       519     &       - -       &       9.8     &       8.4     &       - -     \\
                12013   &       21.--26.03.2014 &       929     &       289     &       - -       &       8.8     &       3.8     &       - -     \\
                12163   &       09.--16.09.2014 &       968     &       443     &       130     &       9.9     &       10.3    &       4.1     \\
                12169   &       18.--26.09.2014 &       1051    &       318     &       266     &       15.2    &       6.8     &       6.5     \\
                12170   &       19.--26.09.2014 &       916     &       356     &       309     &       12.6    &       6.8     &       6.7     \\\hline\hline
                {\bf mean}&                                     &{\bf 881} &{\bf 398}      &{\bf 251}      &{\bf 11.9} &{\bf 8.4}  &{\bf 6.9}      \\\hline
                \end{tabular}}
                \tablefoot{$\bar \upsilon_\mathrm{h}$(3\,pixel) is the mean maximum horizontal velocity (average over the first day of analysis), which is found at the sunspot boundary. $\bar \upsilon^{\mathrm{max}}_\mathrm{h}(r_{\mathrm{F}})$ is the mean maximal velocity after the penumbra/the spot has fully dissolved for different locations in the flow field.}
        \label{tab:list_sunspots}
\end{table*}
%%%%%%%%%%%  TABLE 2 %%%%%%%%%%%%%%%
%                                Horizontal flow velocity - With penumbra
\subsubsection{Horizontal flow velocity}
\begin{figure*}[!h]
\centering 
\includegraphics[width=1.\textwidth]{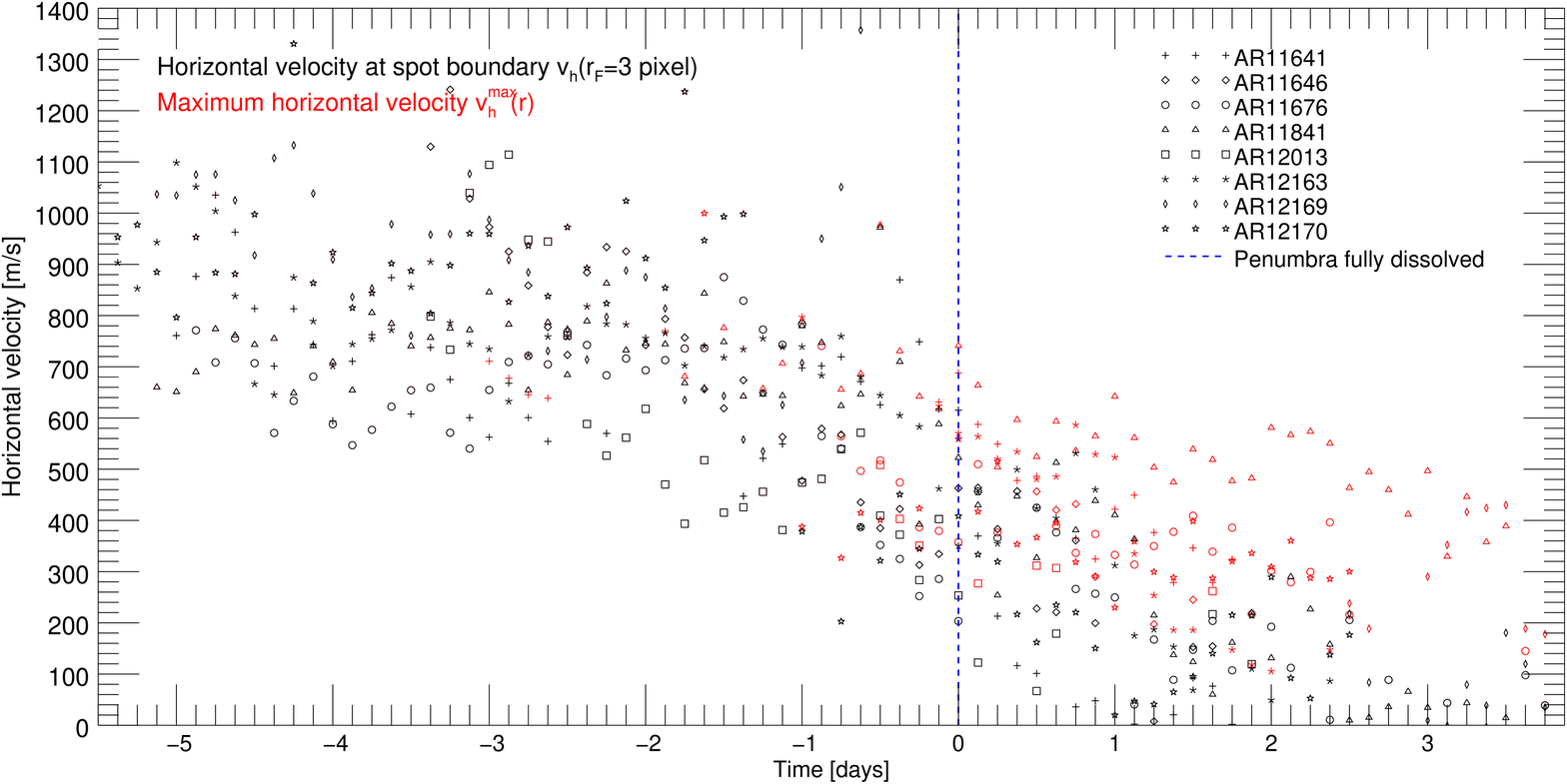}
\caption{Evolution of the maximum velocity $\upsilon^{\mathrm{max}}_\mathrm{h}(r_\mathrm{F})$ (black) and velocity at the spot boundary $\upsilon_\mathrm{h}$(1.1\,Mm) (red) of the flow field for all analysed active regions. If the maximum velocity is localised at the sunspot boundary, these values are equal to $\upsilon_\mathrm{h}$(1.1\,Mm) and are not visible in the diagram. The abscissa values are given in days with the loss of the penumbra being set as a common point in time.} 
\label{fig:velevolution}
\end{figure*}

At their initial stage in the analysis, that is, during their fully fledged evolutionary stage, all sunspots show a velocity profile that decreases continuously from a maximum value at 1.1\,Mm ($r_\mathrm{F}=3$\,pix) off the spot boundary with increasing distance to the spot boundary. See an example for active region (AR) 11641 in Fig.\,\ref{fig:velocityprofile}. On the first day of the analysis, the measured velocities range from a mean maximum velocity (average over the eight-sunspot sample) of 881\,m\,s$^{-1}$ at a distance of 3\,pixels from the boundary of the sunspot (see Table\,\ref{tab:list_sunspots}, third column) down to 199\,m\,s$^{-1}$ at the outer edge of the flow region.
%                                Horizontal flow velocity - Dissolving of penumbra
During the decay of the sunspots, the maximum velocity values show individual changes due to the distinct evolution of each sunspot. Nevertheless, all studied sunspots show a decrease in velocities at the inner boundary of the flow field during sunspot decay. This evolution of the velocity measured closest (1.1\,Mm off) to the sunspot is illustrated in Fig.\,\ref{fig:velevolution} (black), with individual symbols for each sunspot. All times are shifted to show the loss of penumbra as a common point in time (t\,=\,0). 

%%%%%%%%%%%  FIGURE 2 %%%%%%%%%%%%%%%
\begin{figure}[!h]
\centering 
\includegraphics[width=0.5\textwidth]{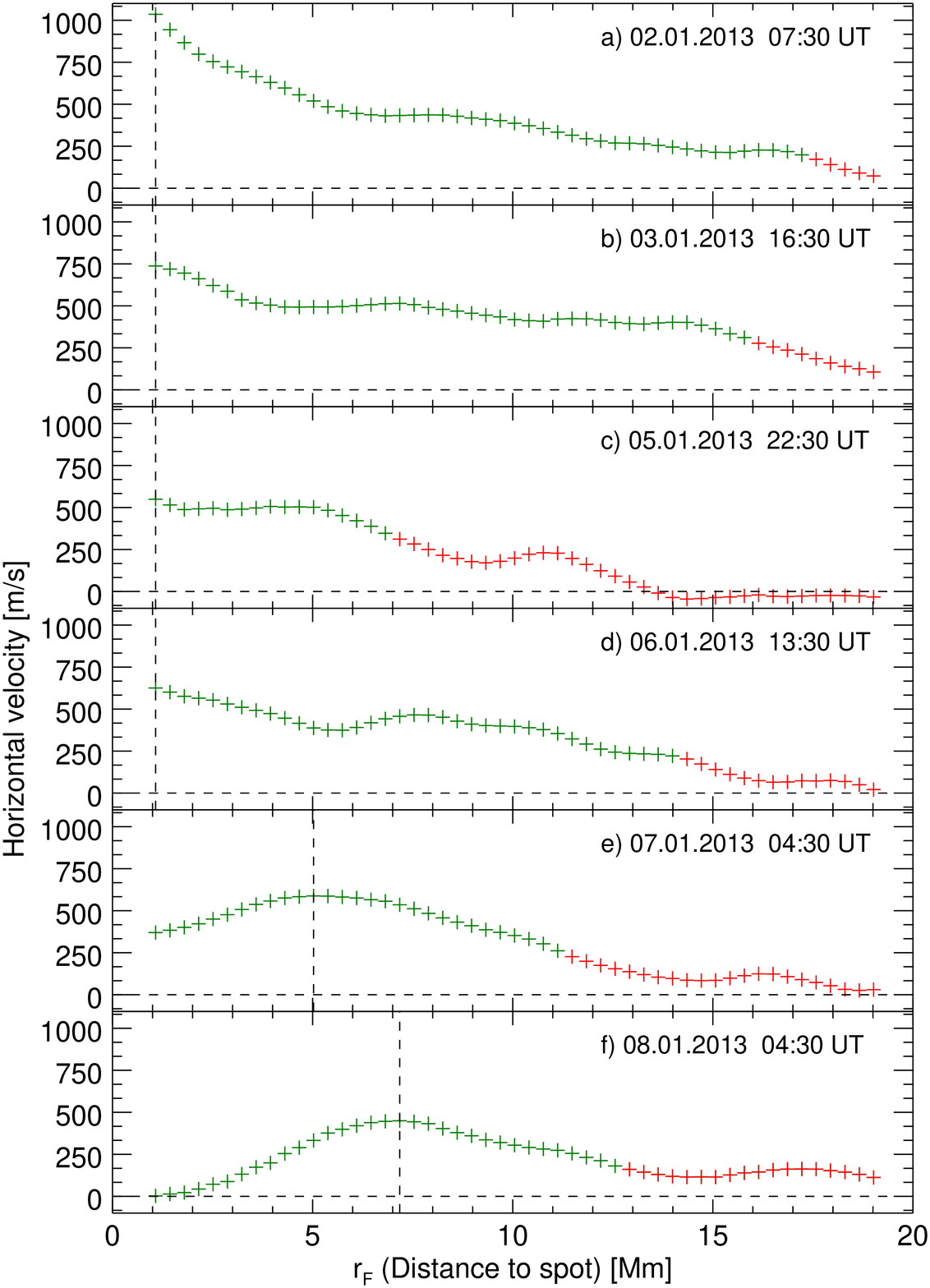}
\caption{Velocity profiles of AR11641 for different times in the spot evolution. Transition from {\em green} to {\em red} in the profiles outlines the outer edge of flow field. A change in the shape of the velocity profile becomes apparent as the sunspot decays and the penumbra dissolves on 7 January 2013 at 01:30\,UT. The vertical lines mark the distance to the spot for which the horizontal velocity is maximum.}
\label{fig:flowprofiles}
\end{figure}

%%%%%%%%%%%  FIGURE 2 %%%%%%%%%%%%%%%
Studying the velocity profiles for the different evolutionary stages of sunspots decay unveils a change in the profile behaviour. The case for AR11641 is illustrated in Fig.\,\ref{fig:flowprofiles}: The horizontal velocity decreases with sunspot decay, especially close to the spot boundary. The continuously decreasing profile, characteristic for the moat flows, disappears. The profile becomes flatter. After the penumbra dissolves (see Fig.\,\ref{fig:flowprofiles}e), the velocity profile changes dramatically: the velocity at the (naked) spot boundary is now lower than the velocity measured 5$-$6\,Mm apart. We estimate a mean maximum horizontal flow of 398\,m\,s$^{-1}$ in the surroundings for the eight\,naked sunspots (see Table\,\ref{tab:list_sunspots}, fourth column, for the individual values of the single spots). 
We note that when determining the average maximum horizontal velocity $\upsilon^{\mathrm{max}}_\mathrm{h}(r_\mathrm{F})$ in the flow field, the distance to the spot boundary is not a fixed value, but rather, $r_\mathrm{F}$ varies for each (naked) spot. This is outlined by the green contour in the animations (see Appendix\,\ref{app:spots}) on the spot evolution.

%%%%%%%%%%%  FIGURE 3 %%%%%%%%%%%%%%%
%%%%%%%%%%%  FIGURE 3 %%%%%%%%%%%%%%%
The change in the shape of the velocity profile becomes even more pronounced as the remaining naked spot dissolves: A velocity profile develops whose amplitude first increases with distance to the (dissolved-)spot centre. A maximum horizontal velocity of 251\,m\,s$^{-1}$ (averaged over the eight ARs) is reached. Farther away from the (dissolved-)spot centre, the velocity decreases again (see e.g. Fig.\,\ref{fig:flowprofiles}f). This mean maximum velocity of 251\,m\,s$^{-1}$ of the flow field measured after the sunspots disappearance is lower than the velocity measured in the surroundings of the naked spots (398\,m\,s$^{-1}$) described above. The two blank rows in Table\,\ref{tab:list_sunspots}, fifth column, are due to measuring maximum velocities after the sunspots disappearance in only six of the eight studied regions. AR11841 does not completely dissolve before reaching the western limb. In the surrounding region of AR12013, no consistent outflow pattern can be measured after the spot has fully dissolved. The behaviour of AR12013 and its surrounding flow field is discussed further in Sect.\,\ref{sec:interact} (where we explain the absence of such an outflow).

%                                Horizontal flow velocity - Maximal and at spot boundary (fig)
To visualise the change in velocity profiles for all active regions, the horizontal velocity by the spot boundary (black) is plotted together with the maximum horizontal velocity (red) measured over the flow field in Fig.\,\ref{fig:velevolution}. As long as the two values coincide, only the values closest to the sunspot are visualised (black). Thus, the detachment of the red and black values reveals the change in the profile of the horizontal velocity component of the radial flow. The onset of the profile change is coincident with the full disappearance of penumbrae, indicated by the blue vertical line in Fig.\,\ref{fig:velevolution}, which shows the intimate coupling between penumbrae and outflows in the surroundings of sunspots.

It is important to note here that the process of decay, distinct for each sunspot, also plays a role in the measured outflows. Thus, a change in the location of the maximum velocity in the flow field before the penumbra has fully dissolved can be ascribed to the separation of part of the penumbra (flux expulsion) from the sunspot, for instance: the evolution of AR11641 (Appendix\,\ref{app:spots}, from 04.01.2013 00:00\,UT -- 09:00\,UT) three days before the penumbra dissolves (red crosses in Fig.\,\ref{fig:velevolution} at t\,=\,-3\,days) shows that the flux expelled into the moat region  influences the measurement of the flow velocity until it vanishes. Similarly, the early onset of the change in velocity profile in AR11841 (red triangles) is due to an asymmetric disappearance of the penumbra (see animation in Appendix\,\ref{app:spots}). The process of the final loss of the penumbra leads to asymmetries in the shape of the other spots as well and therefore also in the surrounding flow fields. This causes the scattering of the shift of the measured locations for the maximum velocity. In addition, the final dissolving of the penumbra is a mostly continuous process, and this point in time is selected by eye in intensity maps, which can also lead to discrepancies and to the spreading around the defined point in time. The reshaping of the radial profile is observed in all cases, however.
%                                Flow extension - difficulty of method
\subsubsection{Flow extension and projection effects}\label{flow_extention}
\begin{figure*}[!h]
\centering 
\includegraphics[width=1\textwidth]{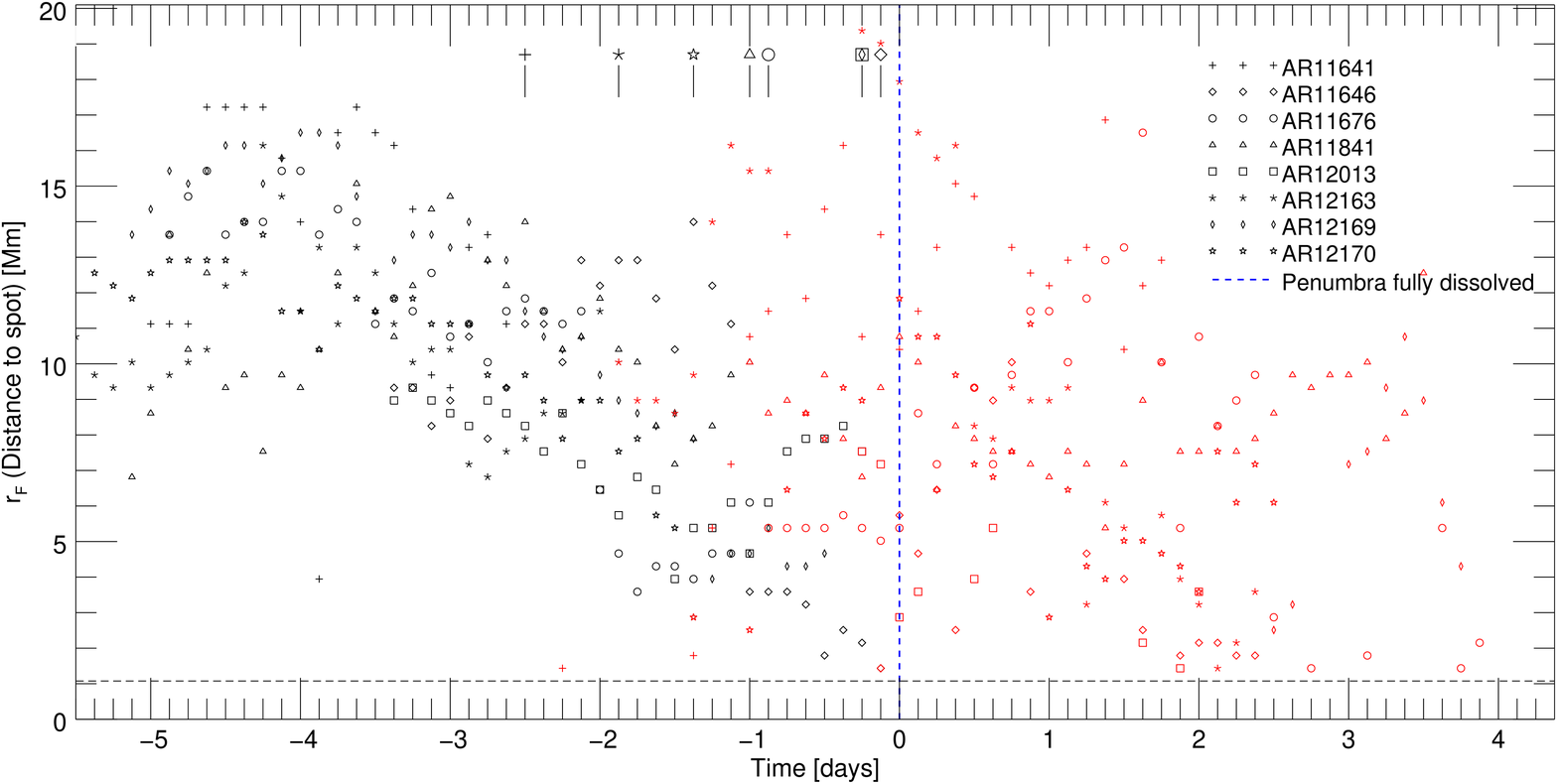}
\caption{Extension of the flow field around all analysed active regions. Black symbols indicate the active regions to be localised on the eastern side on the meridian, while red symbols show values for the active region in the western hemisphere. The black vertical lines and symbols display the approximate crossing of the meridian by the respective sunspot in time. The loss of the penumbra is set as a common point in time.} 
\label{fig:extensionflow}
\end{figure*}

The analysis method of the flow extension as described in Sect.\,\ref{sec:anaflow} relies on the fact that off disc centre, and due to projection effects, a horizontal flow shows a component along the LOS with measurable velocities. Therefore, the measured flow extensions depend on the LOS horizontal velocity and on the $rms$ values. Closer to disc centre, the projection effect decreases, and lower velocity values are measured. The lower values vanish easily in the noise and lead to smaller measured flow extensions close to disc centre. This effect is shown in Fig.\,\ref{fig:extensionflow} for the individual sunspots, as the extension of the flow field seems to decrease and at some point in time starts to increase again when sunspots cross over the zero-meridian. Velocities of sunspots in the eastern and western hemisphere are outlined by black and red symbols. The values are again shifted to have the loss of the penumbra as a common point in time. We note that the analysed time range is different for each sunspot and does not start at the same point in time in the sunspot evolution. Sunspot decay, especially the loss of the penumbra, leads to a large scattering in the values of the flow extension. The erosion and detachment of magnetic flux from the decaying sunspot also affects the measurement of the extension of the outflow.

%                                Flow extension - evolution
For fully fledged sunspots, a mean flow extension of 11.9\,Mm is measured (see Table\,\ref{tab:list_sunspots}, sixth column, for the mean value for each spot). Figure\,\ref{fig:extensionflow} displays a decrease in the flow extension as evolution progresses when comparing similar heliocentric angles on the eastern (black) and western (red) hemispheres of the solar disc. After the penumbra has dissolved, the extension of the outflow also decreases to a mean value of 8.4\,Mm. The values for each sunspot before and after losing the penumbra, as given in Table\,\ref{tab:list_sunspots} (sixth and seventh column) display an individual behaviour of the flow extension for the spots. Around some active regions, for example, AR11641 and AR11841, only a slight decrease is found. For AR12163, even a higher value is measured. This can be ascribed to the influence of a part of the penumbra detaching from the spot. However, the extension of the flow field around other active regions, such as AR12169 and AR12170, decreases by more than 5\,Mm.

We also measure the extension of the flow after the naked sunspot has dissolved. We find a mean extension of 6.9\,Mm for the remaining flow cell without spot. Nevertheless, a distinct behaviour for each flow cell is again obtained (see Table\,\ref{tab:list_sunspots}, last column). The flow cell size of AR11641 remains mostly stable when the values of the different steps of the sunspots evolution are taken into account. The flow extension around AR12169 and AR12170, for instance, also remains constant from the stage of naked spot to the eventual spot extinction. The flow cell around some other active regions, such as AR11646 and AR12163, decreases within half a day after the spot disappears, however. The flow field around AR12013 even disappears immediately with the disappearance of the spot, as we discussed earlier.
%                       Magnetic network - with penumbra
\subsubsection{Interaction between the flow field around sunspots and surrounding network (supergranular) cells}\label{sec:interact}
\begin{figure}[!h]
\centering 
\includegraphics[width=0.5\textwidth]{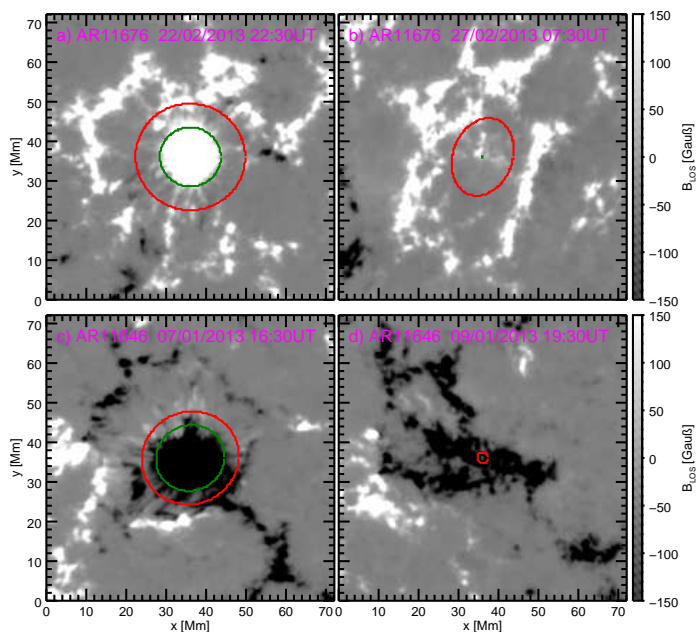}
\caption{LOS magnetograms of AR11676 (upper panels) and AR11646 (lower panels) for different stages in their evolution (left columns: sunspots with penumbra, and right columns: dissolved spots). The area of the spot is outlined by the green ellipses. The extension of the flow field is outlined by the red ellipses.} 
\label{fig:magn}
\end{figure}

We find a main common behaviour for the horizontal flow velocity profile of the different sunspots, but the extension of the flow field surrounding a sunspot shows a more individual behaviour. \citet{zwaan1992} observed breaks in the flow field around a sunspot and suggested a dependence of the shape of the cell on the magnetic environment. To study whether the surrounding of sunspots influences the extension of the flow field, we used HMI LOS magnetograms. When sunspots have a penumbra, the LOS magnetograms show the well-known scenario of a sunspot as a strong accumulation of magnetic flux with moving magnetic features streaming away from it, and with the spot centred in a (network) cell, as shown in Fig.\,\ref{fig:magn}a. The magnetic flux accumulating around the spot seems to form a firm boundary against the surrounding network cells. The network (supergranular) cells evolve in continuous competition with each other, including the sunspot cell. At some locations, the surrounding network cells are able to interrupt the sunspot's flow cell, as is shown in the network cell centred at around x\,=\,56\,Mm, y\,=\,26\,Mm, in Fig.\,\ref{fig:magn}a, for instance. In other cases, such as Fig.\,\ref{fig:magn}c, the sunspot cell implodes, squeezed by the surrounding cells. Nevertheless, as long as the penumbra is present, the cell formed around the sunspot is only partially distorted and can re-establish itself again and push the surrounding network cells back. The evolution of each sunspot as seen in the magnetograms can be followed in the central panel of the animations (see Appendix\,\ref{app:spots} and Figs\,\ref{fig:ar1_fig} - \ref{fig:ar8_fig} for snapshots).
%%%%%%%%%%%  FIGURE 4 %%%%%%%%%%%%%%%
%%%%%%%%%%%  FIGURE 4 %%%%%%%%%%%%%%%
%%%%%%%%%%%  FIGURE 5 %%%%%%%%%%%%%%%
%%%%%%%%%%%  FIGURE 5 %%%%%%%%%%%%%%%
%                       Magnetic network - dissolve of penumbra
As the penumbra dissolves, the morphology of the spot changes and the strength of the flow seems to weaken. This influences the measured extension of the flow field (see previous Sect.\,\ref{flow_extention}). We observe two different scenarios: 
%                       Magnetic network - flow cell remains
(1) For some cases, in regions of the spot cell boundary in which magnetic flux accumulates, the influence of surrounding (supergranular/network) cells appears to be weak and a cell can form around the spot. In such cases, the cell remains stable, even after the spot vanishes, and the flow extension can be measured. An example is illustrated in Fig.\,\ref{fig:magn}b for AR11676. With time, the cell size slowly decreases as the surrounding network cells push against it, leading to a long-lasting (>\,1\,day) dissolving process. The full evolution process can be monitored in the animations in Appendix\,\ref{app:spots}.
%                       Magnetic network - no flow cell
 (2) In contrast, if there is weak magnetic flux in the spot cell boundary, the surrounding cells are able to distort the flow field by squeezing the spot cell. This leads to a decrease in flow extension. The effect of squeezing the spot cell becomes even more pronounced as the spot vanishes in the intensity maps:
Eventually, the spot cell implodes owing to the action of the surrounding cells, as shown in Fig.\,\ref{fig:magn}c and Fig.\,\ref{fig:magn}d for AR11646, for instance. We observe a similar behaviour for AR12013 and AR12163. For the former, the flow even disappears with the disappearance of the spot.

%                       Magnetic network - conclusion
These observations suggest that the variety of measured extension values of the flow field surrounding a spot (Sect.\,\ref{flow_extention}) can be explained by boundary conditions, namely, the interaction with surrounding network cells. In addition, the accumulation of magnetic flux, forming part of the network, seems to influence the stability of the cell.
%===================================================================================
%                                Flow field of supergranules  
\subsection{Flow field of supergranules}\label{sec:sgs}
To analyse the flow field in supergranules, we applied the same method as for the study of the flow field around sunspots. To localise the centre of supergranules, we used LOS magnetograms and assumed the magnetic network as a suitable outliner for the supergranular pattern manifest in the averaged Doppler maps. The centre of such a cell was then manually localised by eye. In this way, we selected 18 supergranules in the magnetograms and read out the LOS velocity values from the Doppler maps along circles and ellipses around the defined barycentre point. Examples of the contouring analysis of three supergranules of the sample as seen in intensity, magnetograms, and Doppler maps can be found in the figures provided in Appendix\,\ref{app:sg}.

%                               Horizontal flow velocity
\subsubsection{Horizontal flow velocity}
The mean horizontal velocity profiles are displayed in Fig.\,\ref{fig:overview}d for all selected supergranules. On average, the velocity increases from a mean value of 215\,m\,s$^{-1}$ for increasing distance to the cells centre, up to a maximum of 355\,m\,s$^{-1}$ to later decrease for larger distances. The maximum horizontal velocity, in the range of 216--511\,m\,s$^{-1}$, is always localised farther out in the flow field and not at the centre of the cell. 

%                                Flow extension
\subsubsection{Flow extension}
According to our analysis method, the cell size is determined as the distance to the cell centre where the measured LOS horizontal velocity is lower than the noise level. We measure a mean distance of 8.8\,Mm, with values in a range from 6.1\,Mm to 12.9\,Mm, which translates into an estimate of the diameter of the cells of 12.2\,Mm to 25.8\,Mm. 

%===================================================================================
%                                DISCUSSION
%===================================================================================
%%%%%%%%%%%  TABLE 3 %%%%%%%%%%%%%%%
\begin{table*}[!ht]
        \caption{Summary of results.}
        \centering 
        \resizebox{1.\linewidth}{!}
        {\begin{tabular}{cccc} \hline
        \rule{0pt}{3pt}\textbf{Spot with penumbra}      &       \textbf{Naked spot}   &       \textbf{Dissolved spot} &       \textbf{Supergranule}   \\\hline
        \rule{0pt}{4pt}$\bar \upsilon^{\mathrm{max}}_\mathrm{h}(r_{\mathrm{F}})=881\mathrm{m\,s}^{-1}$ &       $\bar \upsilon^{\mathrm{max}}_\mathrm{h}(r_{\mathrm{F}})=398\,\mathrm{m\,s}^{-1}$    &       $\bar \upsilon^{\mathrm{max}}_\mathrm{h}(r_{\mathrm{F}})=250\,\mathrm{m\,s}^{-1}$    &       $\bar \upsilon^{\mathrm{max}}_\mathrm{h}(r_{\mathrm{F}})=355\,\mathrm{m\,s}^{-1}$            \\
        \rule{0pt}{3pt}($\bar \upsilon^{\mathrm{max}}_\mathrm{h}(r_{\mathrm{F}})=1000\,\mathrm{m\,s}^{-1}$)\textsuperscript{(1)} & ($\bar \upsilon^{\mathrm{max}}_\mathrm{h}(r_{\mathrm{F}})=500\,\mathrm{m\,s}^{-1}$)\textsuperscript{(3)}& &($\bar \upsilon^{\mathrm{max}}_\mathrm{h}(r_{\mathrm{F}})=300-500\,\mathrm{m\,s}^{-1}$)\textsuperscript{(2,5)}\\\hline
        \rule{0pt}{3pt}$\bar r_{\mathrm{F_{end}}}=11.9\,\mathrm{Mm}$    &        $\bar r_{\mathrm{F_{end}}}=8.4\,\mathrm{Mm}$    &       $\bar r_{\mathrm{F_{end}}}=6.9\,\mathrm{Mm}$    &       $\bar r_{\mathrm{F_{end}}}=8.8\,\mathrm{Mm}$  \\
        \rule{0pt}{3pt}($\bar r_{\mathrm{F_{end}}}=9.2\,\mathrm{Mm}$)\textsuperscript{(1)}         &        ($\bar r_{\mathrm{F_{end}}}=10\,\mathrm{Mm}$)\textsuperscript{(3)}     &               &       $(\bar r_{\mathrm{F_{end}}}=5-22.5\,\mathrm{Mm}$)\textsuperscript{(2,4)}       \\\hline
        \end{tabular}}
         \tablefoot{ The first three columns display the mean values obtained for the different stages of the sunspots evolution (sunspot with penumbra, naked spot, dissolved spot). The fourth column depicts the mean values of the supergranular flow.}
        \tablebib{(1)~\citet{loehner-boettcher2013}, (2)~\cite{orozcosuarez2012}, (3)~\citet{rempel2015}, (4)~\citet{roudier2014}, (5)~\citet{svanda2014}}
        \label{tab:overview}
\end{table*}
%%%%%%%%%%%  TABLE 3 %%%%%%%%%%%%%%%
\section{Discussion}\label{sec:disc_conc}
We have studied how the radial outflow around sunspots, the moat flow, evolves with sunspot decay. For comparison, we have measured the flow in supergranules by similar means. In the following, we discuss our results in the context of the findings reported by other authors, and we investigate the possible relation between these two large-scale plasma flows.
%Flow around sunspots - moat flow with penumbra
\subsection{Flow around fully fledged sunspots}
The measured extension of the moat flow around fully fledged sunspots (day one in our evolutionary study) are in agreement with the results obtained by \citet{loehner-boettcher2013}, who applied the same analysis method. Our mean extension value of 11.9\,Mm (see Table\,\ref{tab:overview}, first column and Fig.\,\ref{fig:overview}a, change from green to red) is somewhat higher than their mean flow extension of 9.2\,Mm. This can be ascribed to our reduced (eight sunspots) sample compared to theirs (31 sunspots). It is well within their range of values of 5--15\,Mm, however. Other investigations by \citet{sheeley1972}, \citet{brickhouse1988}, \citet{balthasar2010}, \citet{svanda2014} and \mbox{\citet{verma2018}}, for example, led to a larger flow extension ranging from 10--22\,Mm. The measured profile shape of the horizontal velocity, showing a continuous decrease with increasing distance from the spot boundary (see Fig.\,\ref{fig:overview}a), agrees with the results by \citet{balthasar2013}, \citet{loehner-boettcher2013}, and \citet{rempel2015}. The maximum velocity values, ranging between 677\,m\,s$^{-1}$ and 1051\,m\,s$^{-1}$  and measured close to the sunspot boundary (grey shaded area in Fig.\,\ref{fig:overview}a at 1.1\,Mm), are comparable to those reported by \citet{loehner-boettcher2013} (0.8 to 1.2\,km\,s$^{-1}$). 
%Flow around sunspots - w/o penumbra
\subsection{Flow around naked sunspots}
As the penumbra dissolves, we measure a decrease in the horizontal velocity around the spot. This is in agreement with the findings by \citet{balthasar2013}, who reported on lower flow velocity on the side of the sunspot where its penumbra has dissolved compared to the outflow velocity where the penumbra is still present. The measured evolution of the velocity profile (see Fig.\,\ref{fig:overview}b) is also in agreement with the results of the simulations by \citet{rempel2015}. He reported a horizontal velocity at a distance of 1\,Mm to the spot boundary that increased with distance up to 500\,m\,s$^{-1}$ to decrease later.
We measure a mean maximum velocity value of 398\,m\,s$^{-1}$ (see Table\,\ref{tab:overview}, second column and Fig.\,\ref{fig:overview}b,) with individual values ranging from 289\,m\,s$^{-1}$ to 519\,m\,s$^{-1}$. In Fig.\,\ref{fig:overview}b, the grey shaded area reaches an even higher value because this area displays the dispersion of all individual profiles for all times during the naked-spot phase and for all the spots in the sample. In addition, the maximum horizontal velocity is not found to be at the same distance to the spot boundary for the different time steps and sunspots. The extension of the flow field reported by \citet{rempel2015} does not change in his simulated naked spot. This disagrees with our measured mean value, which decreases from 11.9\,Mm to 8.4\,Mm (compare Table\,\ref{tab:list_sunspots}, Cols.\,6 and 7 and Table\,\ref{tab:overview}, Cols.\,1 and 2), although we also have active regions that show a more constant flow extension (e.g. AR11646, AR11841, and AR12163, see Table\,\ref{tab:list_sunspots}) during sunspot decay. According to our study, the interaction with surrounding cells (see Sect.\,\ref{sec:interact}) causes the measured decrease of the flow extension. The discrepancy of the simulations by \citet{rempel2015} with our results can be ascribed to the missing influence of supergranular flows in the surrounding of the naked-sunspot cell in the simulations. This was also noted by \citet{sheeley2017}, whose observations and suggestion of supergranules compressing the outflow on one side agree with our observations and interpretations. \citet{sheeley1972} and \citet{zwaan1992} have also reported about irregularities in the flow cell around sunspots due to sections of the network penetrating it. In addition, \citet{meyer1974} explained discrepancies in the measurement values of decaying sunspots as the lack of symmetry and incompleteness of the flow field. In our analysis, the horizontal velocity is obtained as the amplitude of the LOS velocity read out along circles around the sunspot (see Sect.\,\ref{sec:anaflow}). The penetration of a neighbouring supergranule into the flow region of the sunspot in the direction to its axis of symmetry will have a stronger influence on the measured value than the penetration of a supergranule perpendicular to the spot axis of symmetry. Therefore, we stress that for an irregular cell shape, the measurement of the flow extension around a naked sunspot, with our analysis method, will depend on the direction of the penetration of the neighbouring supergranules.
\subsection{Flow in the remnant region of dissolved spots}
\begin{figure}[!ht]
\centering
\includegraphics[width=0.5\textwidth]{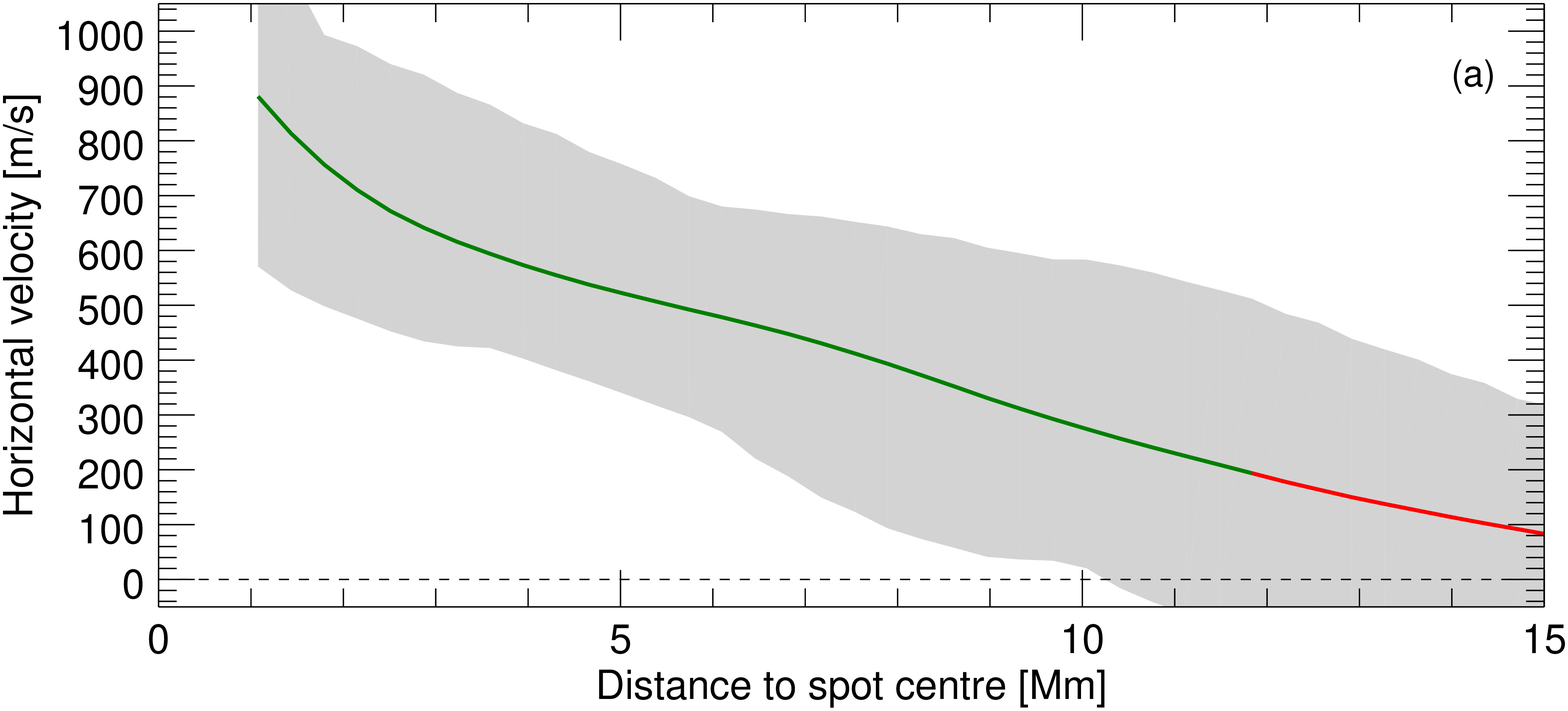}
\includegraphics[width=0.5\textwidth]{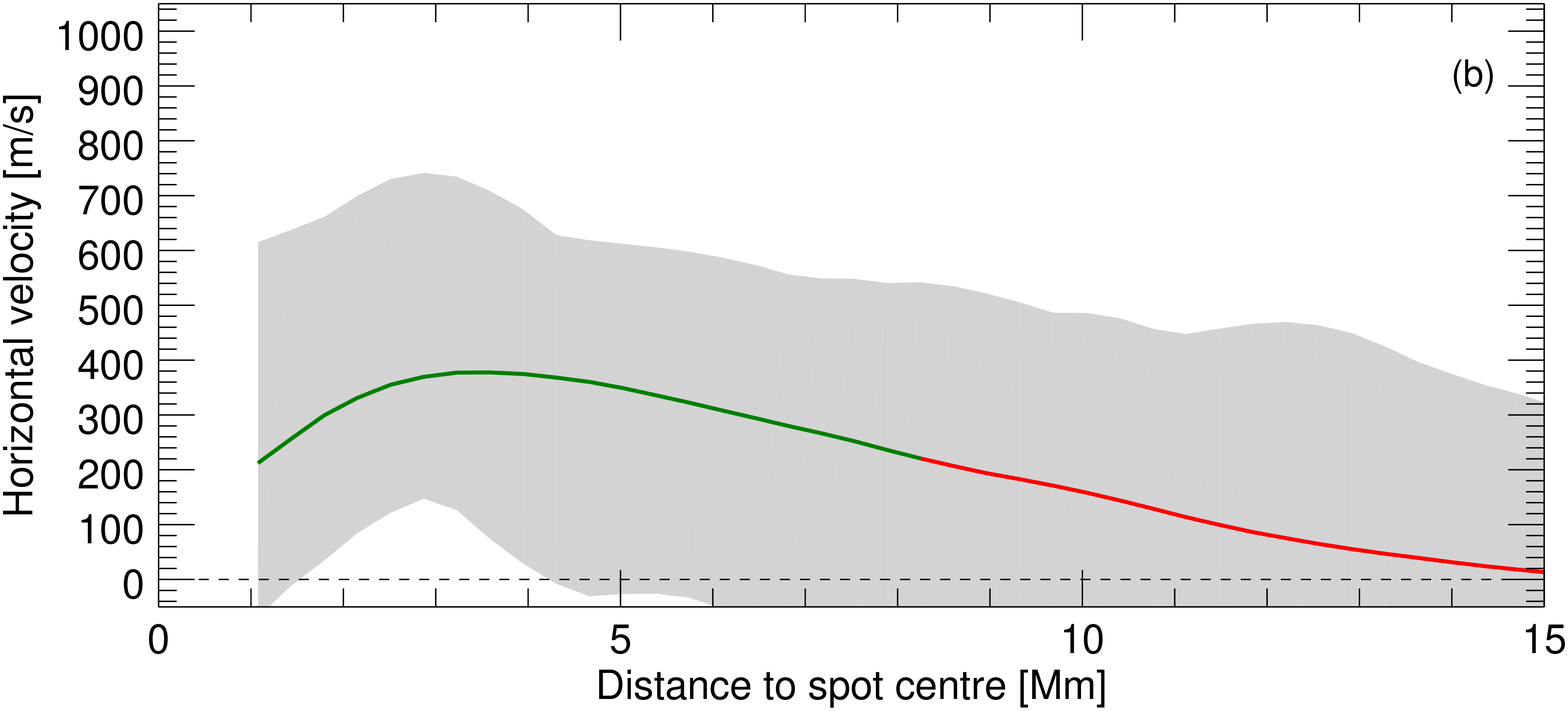}
\includegraphics[width=0.5\textwidth]{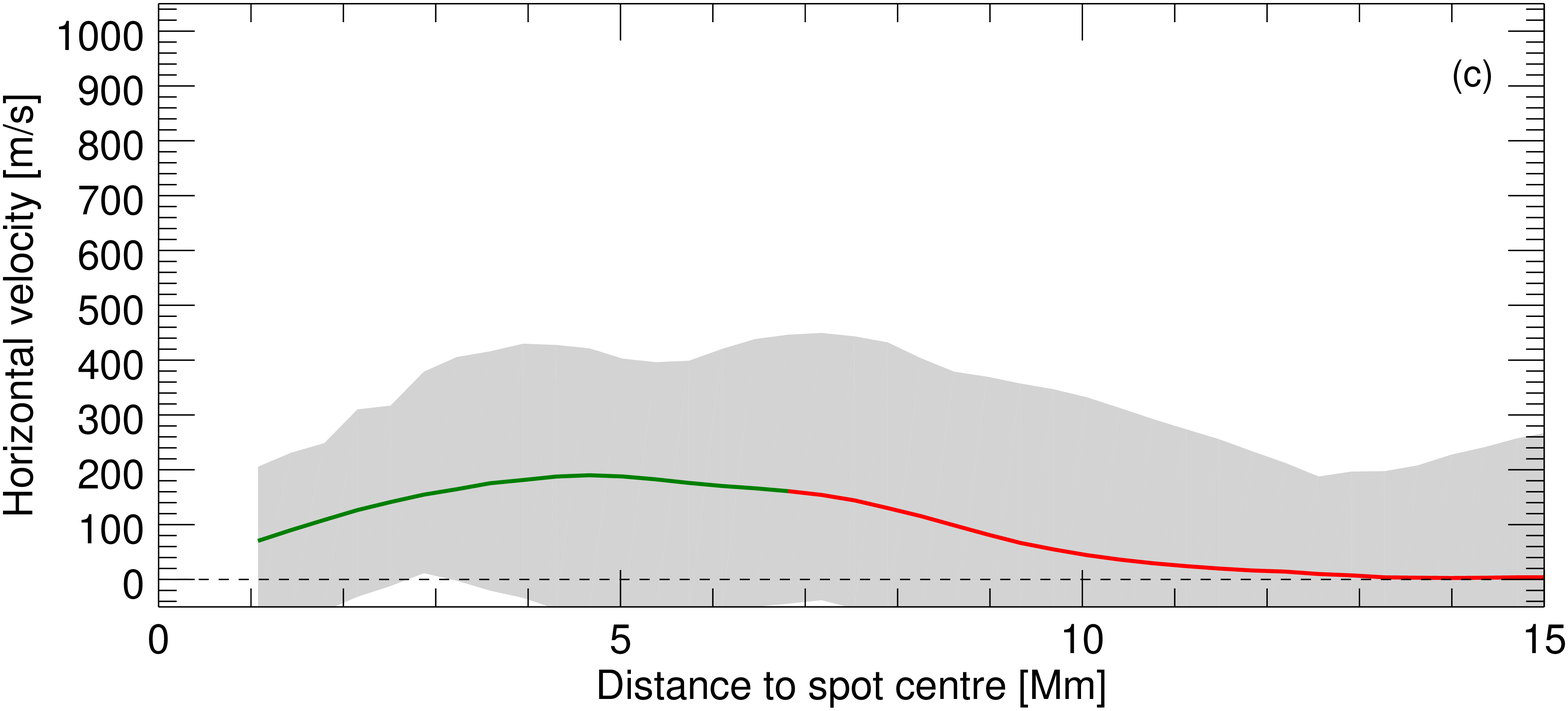}
\includegraphics[width=0.5\textwidth]{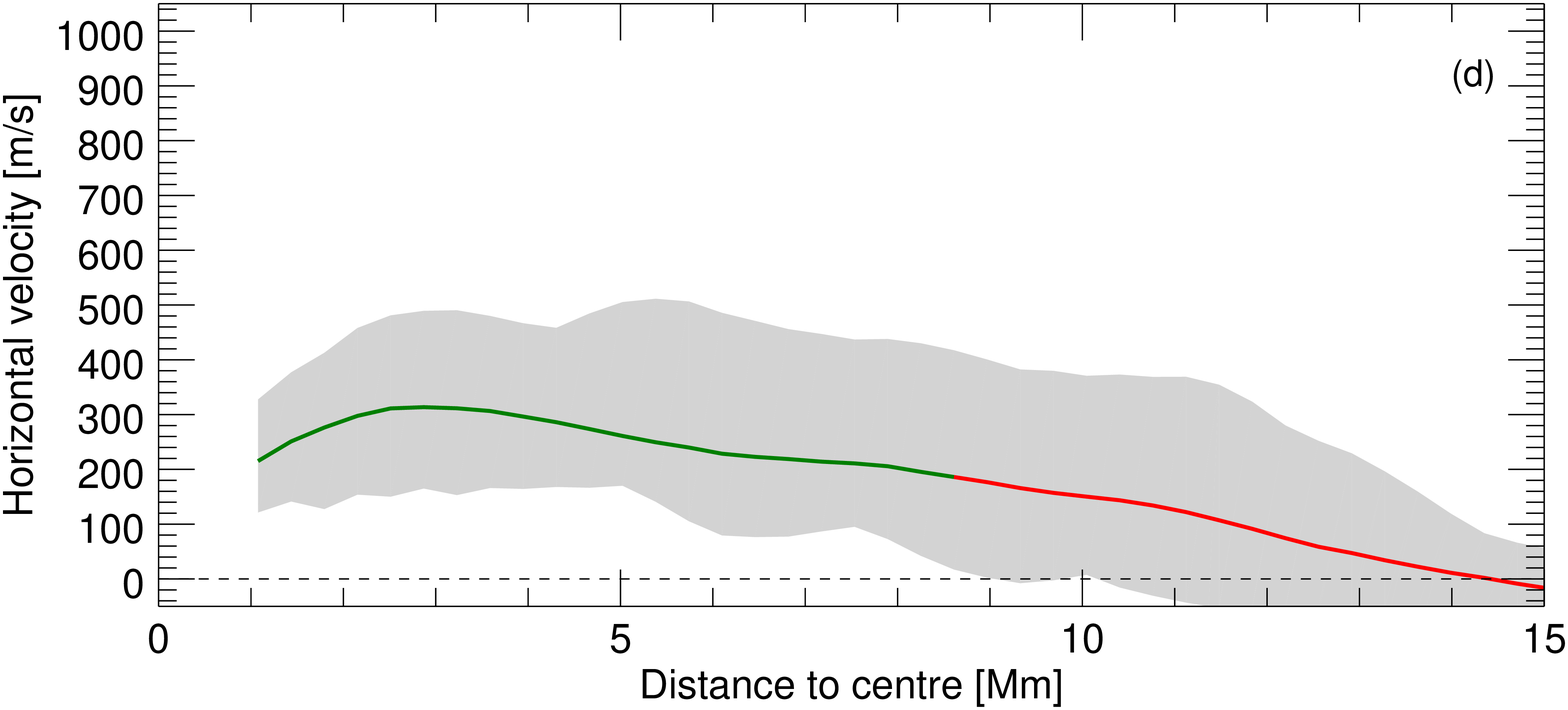}
\caption{Mean horizontal velocity of the flow field of (a) fully fledged sunspots, (b) naked spots, (c) dissolved spots, and (d) supergranules.}
        \label{fig:overview}
\end{figure}

For the six analysed cases in which a flow can be measured after the spot has dissolved, we measure a mean extension of 6.9\,Mm (see Table\,\ref{tab:overview}, third column and Fig.\,\ref{fig:overview}c, change from green to red) and a mean maximum horizontal velocity of 250\,m\,s$^{-1}$. This means that after sunspot decay, the remaining flow cells have a mean diameter of 13.8\,Mm. These cells are mostly outlined by the magnetic network as seen in magnetograms (see animations in Appendix\,\ref{app:spots} and Figs.\,\ref{fig:ar1_fig} - \ref{fig:ar8_fig}, centre panels). Directly after the disappearance of the spot as seen in the intensity maps, it can be distinguished from a regular network cell (supergranule) because of the enhanced magnetic flux remaining from the spot. This flux is then spread in time, possibly advected by convection as described by \citet{vandriel-gesztelyi2015}. To interpret the remaining velocity profile (Fig.\,\ref{fig:overview}c), we recall that the grey shaded area takes into account all individual profiles obtained for the different remnant cells and the maximum values are not localised at the same distance to the cells' centre. Therefore, the individual profiles would show intersections, that is, some might reach maximum velocity closer to the centre and show a faster decreasing radial velocity profile than others.

\citet{pardon1979} reported on 2 (out of 18) moat flows that persisted for two days after the spot disappeared, as seen in magnetograms. We are not aware of any other study on the remnant flow cell of a decaying sunspot. Because of the evident similarities between supergranules and these remnant cells as seen in magnetograms and Doppler maps, we compare the results we obtained for both of them in Sect.\,\ref{sec:sg-flow}.
%%%%%%%%%%% FIGURE 6 %%%%%%%%%%%%%%%
%%%%%%%%%%% FIGURE 6 %%%%%%%%%%%%%%%
%Supergranular flow
\subsection{Supergranular flow}
Our study of the flow in 14 supergranules led to a mean diameter of 17.6\,Mm for supergranular cells (see Table\,\ref{tab:overview} last column). This is smaller than the mean cell diameter of 25\,Mm reported by \citet{roudier2014}. Nevertheless, our individual values, in the range of 12.2\,Mm to 25.8\,Mm, are within the diameter of supergranular cells measured by \citet{simon1964} and \citet{roudier2014}, which vary from 10\,Mm to 45\,Mm. We surmise that our lower mean value is due to our limited sample and the manual selection of the single cells. This was done by eye, and therefore cells with approximately the same size have been selected. \cite{hirzberger2008} also ascribes the wide range of cell sizes to the different analysis methods reported by different authors. The horizontal velocity profiles measured for the supergranular flows resemble each other. They all increase with the distance to the cell centre, reach a (averaged over all samples) mean maximum velocity of 355\,m\,s$^{-1}$ , and decrease for larger distances (see Fig.\,\ref{fig:overview}d). Our supergranular velocity profiles share a similar shape and maximum velocity with those by \cite{orozcosuarez2012}, who fitted their profiles to the supergranular convection kinematic model by \citet{simon1989}. Yet, our flow cells are somehow smaller. Our maximum velocities also agree with those measured by \cite{svanda2014}.
%Supergranular flow and flow in decaying active regions - with naked spot
\subsection{Supergranular flow and flow in decaying active regions}\label{sec:sg-flow}
By comparing the radial velocity profile of the supergranular flow and the flow field around sunspots after the penumbra has fully dissolved, the similar curve progression becomes obvious (see Fig.\,\ref{fig:overview}b and d). The mean maximum velocity of the flow around the remaining naked spot of 398\,m\,s$^{-1}$ is somewhat higher than the mean maximum horizontal flow velocity in supergranules (355\,m\,s$^{-1}$) (see Table\,\ref{tab:overview}, Cols.\,2 and 4), but still within the range of single measured values. The mean value of the flow extension around the naked spot of 8.4\,Mm is of the same size as the measured extension of a mean supergranule (8.8\,Mm).

%Supergranular flow and flow in decaying active regions - w/o naked spot
Once the decaying spot dissolves, the remaining cell certainly resembles a network cell in magnetograms. Comparison of the individual cases shows that the remaining flow in the active regions squeezed by the surrounding supergranules do not share similarities with standard supergranular flows (see e.g. AR11646 and AR12013, Table \ref{tab:list_sunspots}). For spots whose cell persists after the spot has dissolved, the remaining flow values (see Table\,\ref{tab:overview}, Col.\,3) agree with the lower range values of supergranular flows, however. This also becomes apparent by comparing the profiles of the horizontal velocity (see Fig.\,\ref{fig:overview}c). The profile for the remnant flow region is flatter, but the maximum velocity is not localised at the spot boundary.
%Supergranular flow and flow in decaying active regions - 
\subsection{Moat flow decay: interaction with the surroundings}
Our results do not allow us to rule out at this point any of the scenarios propsed by \citet{sheeley1972} or \citet{vargasdominguez2008}, as described in Sect.\,\ref{introduction}. We clearly corroborate the intimate relation between the decay of the moat flow around sunspots and the disappearance of the penumbra during active region decay.

The close agreement of the properties measured in the flow around naked sunspots and the flow in supergranules leads us to conjecture that a supergranule takes over the remaining flow cell in a decaying active region. This remaining flow appears to be more stable as long as the naked spot is still present, and it weakens when the spot fully dissolves. The interaction with the surrounding supergranules is observed to result in the expansion and eventual fragmentation of the flowing cell or its merging as a result of implosion, as also described by \citet{simon1968} and \citet{derosa2004}. In this process, the remaining magnetic flux would either be spread over a larger area (\citealt{harvey1973}) by advection or be concentrated, hence becoming part of the network, respectively (see Fig.\,\ref{fig:magn}). 

Our results lead us to conclude that the presence of a flow cell enclosing a spot and how it evolves strongly depends on the behaviour of the surrounding competing supergranules and related magnetic network, that is to say, it strongly depends on the surrounding environment and not only on the presence of the sunspot itself. Our findings are in line with those by \citet{simon1968}, who concluded from their observations that the shape of the moat depends on the magnetic surrounding and ascribed breaks in the moat to substantial magnetic structure nearby.
%===================================================================================
%                                CONCLUSION
%===================================================================================
\section{Summary and conclusions}\label{sec:concl}
From the detailed analysis of the flow velocity, especially the variation in radial velocity profile and the flow extension, the evolution of the flow field around sunspots can be described as follows: For fully fledged sunspots, the flow is characterised by a continuous outwards plasma motion, the moat flow. The horizontal velocity component is maximum close to the sunspot as long as the penumbra is present, and decreases while the sunspot decays. The moat flow generates a moat cell around the spot, which is generally outlined by a magnetic boundary. Network cells (supergranules) outlined by the magnetic network in the surrounding area interact with the moat flow. Expanding network cells push the moat flow backwards, towards the sunspot, thus destroying the annular form of the moat cell. The moat flow is mostly strong enough to reinstall and also repel the competing network (supergranular) cells. This leads to the asymmetric shape of most moat cells.

When the penumbra dissolves, the flow profile changes. As the naked spot remains, the surrounding outflow resembles the flow in supergranules, that is, the maximum velocity is now found at larger distance from the spot boundary. The interaction between this flow field and the surrounding network cells remains. Without penumbra and decreasing flow velocities, however, the neighbouring cells are able to further push back the outward-moving plasma towards the decaying sunspot. This can lead to the deformation of the cell. In some cases, the network cells are even squeezed until they reach the boundary of the naked spot. If the remaining outflow is still strong enough, a cell smaller than the original moat cell prevails, which then dissolves within supergranular timescales.

Early studies by \citet{meyer1974}, \citet{simon1964}, and \citet{sheeley1972} have proposed the possible coupling between the supergranular and moat flows. Our study does not allow for a definite proof of this relation. Our findings, making use of HMI data, give indications of the moat flow transitioning into a supergranular flow when the penumbra disappears or sunspot decays, however, which supports the findings and scenarios proposed by those authors.

The supergranular convection kinematic model by \citet{simon1989} also agrees with our findings for the flow in supergranules. This also holds for the results of \cite{orozcosuarez2012}, although our flow cells are somehow smaller.

\begin{figure*}[!h]
\centering 
\onecolumn\includegraphics*[width=1.\linewidth]{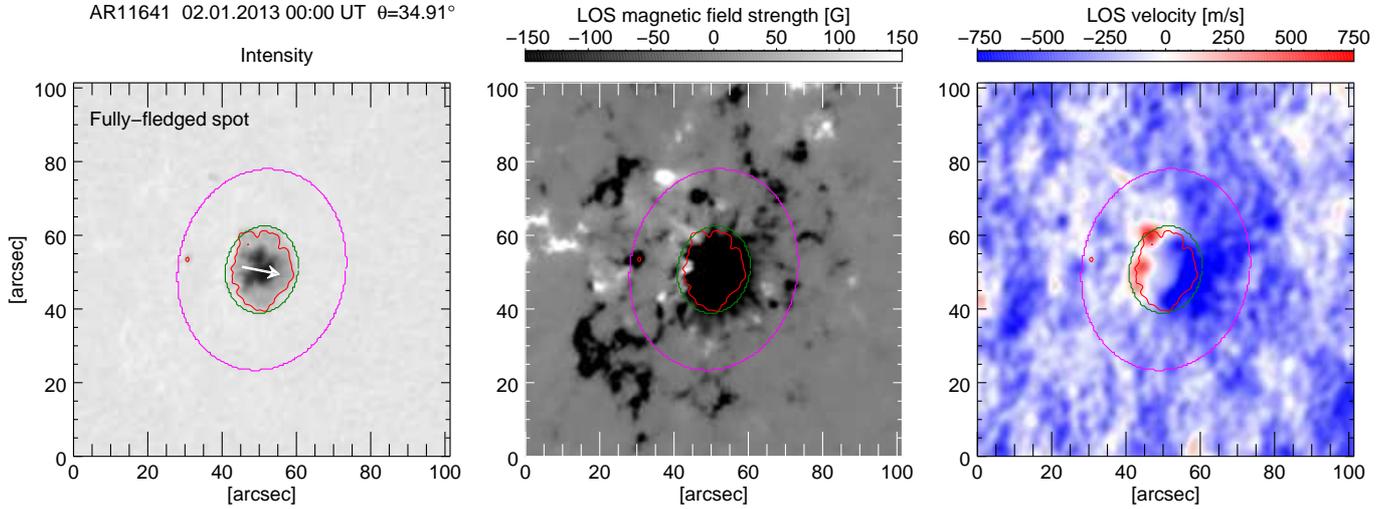}
\caption{Snapshots of the evolution AR11641 in intensity maps (left), LOS magnetograms (middle), and LOS Doppler maps (right). The position of the maximum horizontal velocity and extension of the horizontal flow are marked by green and pink contours, respectively. The white arrow in the intensity map points in the direction of disc centre.} 
\label{fig:ar1_fig}
\end{figure*}

\begin{figure*}[!h]
\centering
\includegraphics[width=1\textwidth]{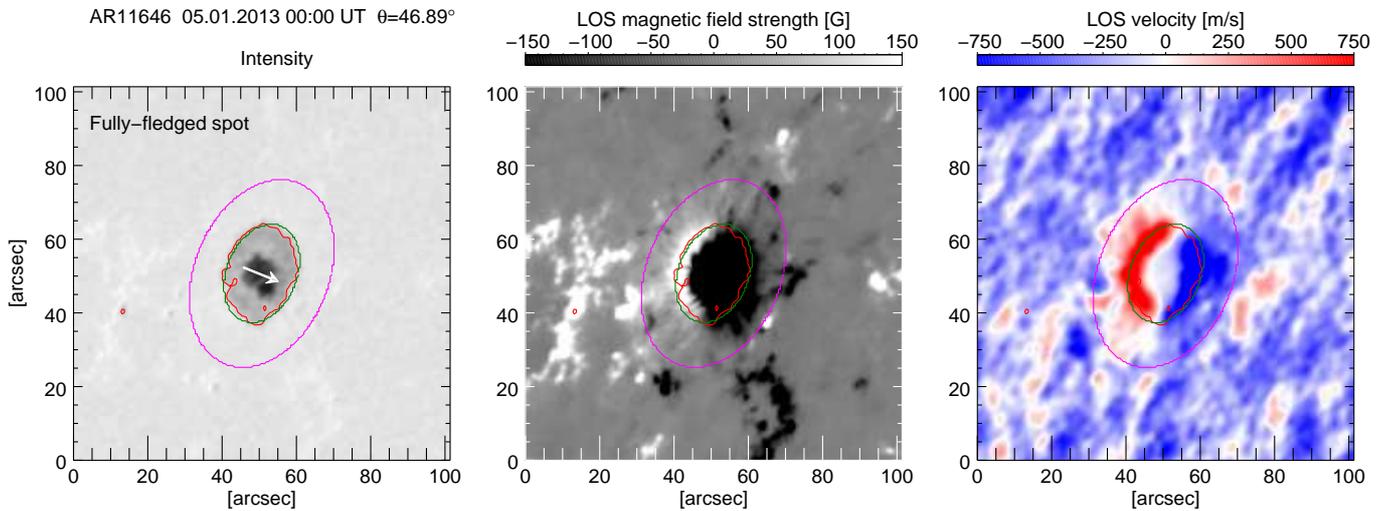}
\caption{Same as Fig.\,\ref{fig:ar1_fig}, except for AR11646.} 
\label{fig:ar2_fig}
\end{figure*}

\begin{figure*}[!h]
\centering
\includegraphics[width=1\textwidth]{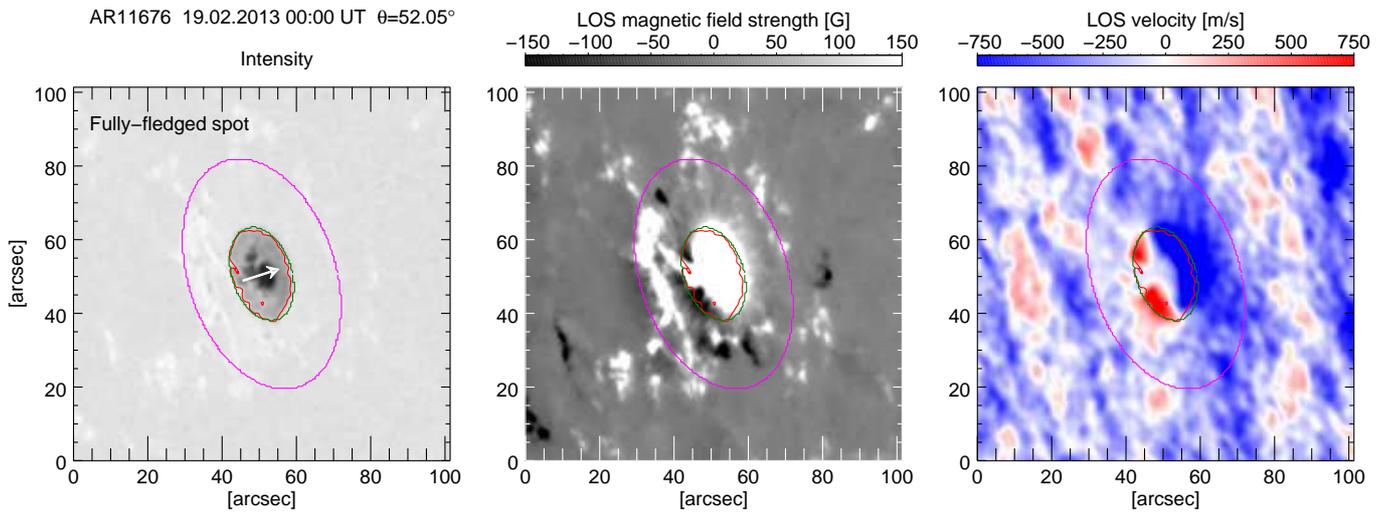}
\caption{Same as Fig.\,\ref{fig:ar1_fig}, except for AR11676.} 
\label{fig:ar3_fig}
\end{figure*}

\begin{figure*}[!h]
\centering
\includegraphics[width=1\textwidth]{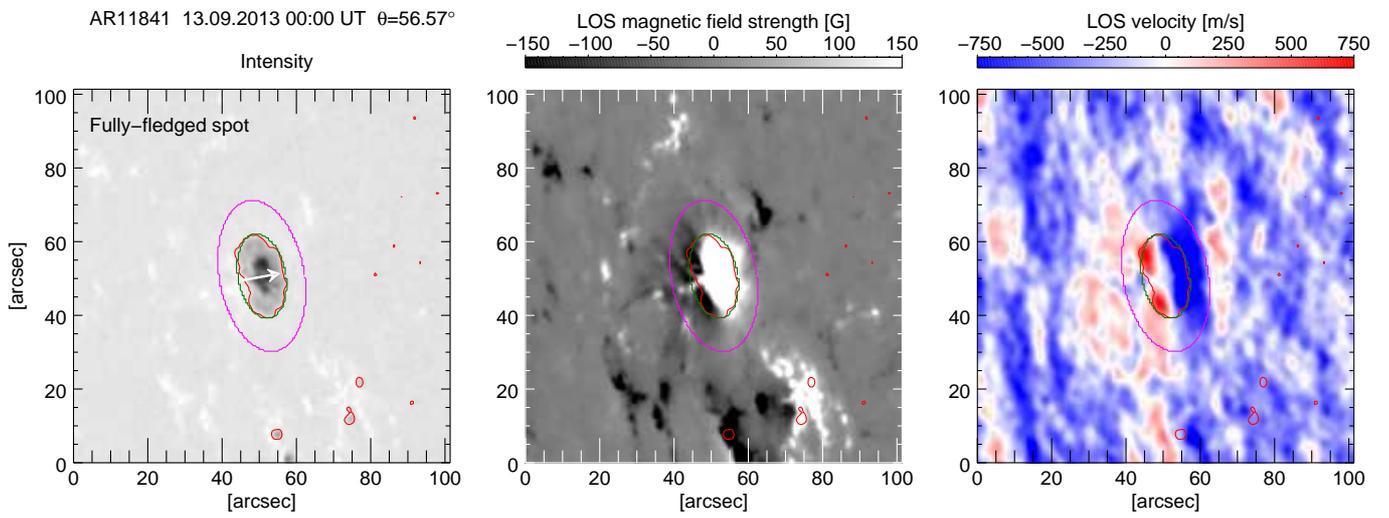}
\caption{Same as Fig.\,\ref{fig:ar1_fig}, except for AR11841.} 
\label{fig:ar4_fig}
\end{figure*}

\begin{figure*}[!h]
\centering
\includegraphics[width=1\textwidth]{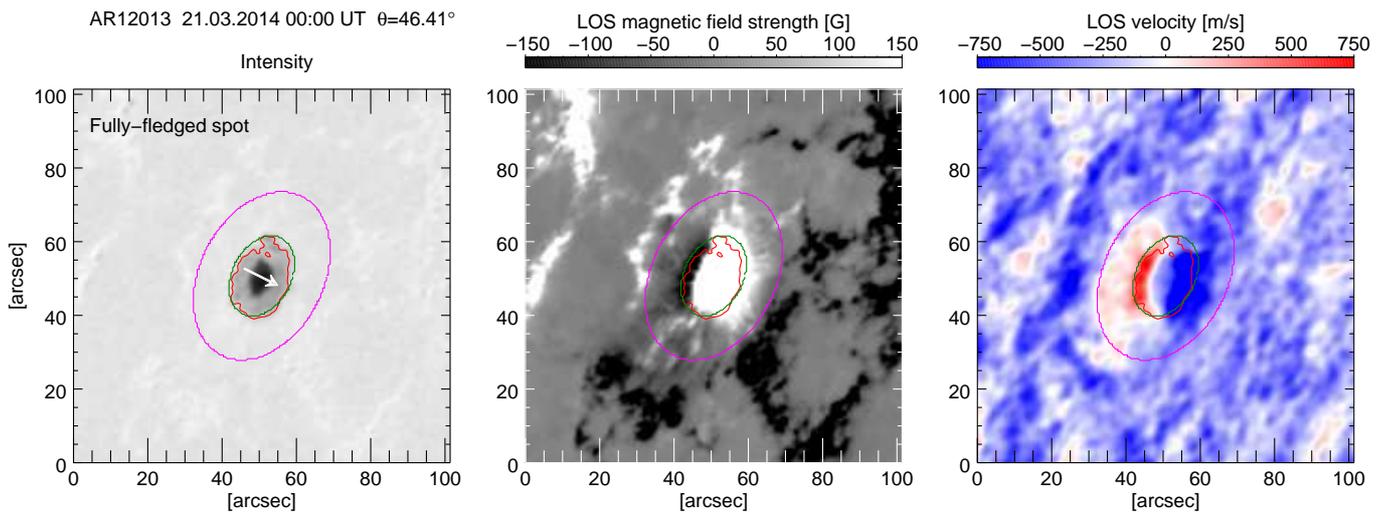}
\caption{Same as Fig.\,\ref{fig:ar1_fig}, except for AR12013.} 
\label{fig:ar5_fig}
\end{figure*}

\begin{figure*}[!h]
\centering
\includegraphics[width=1\textwidth]{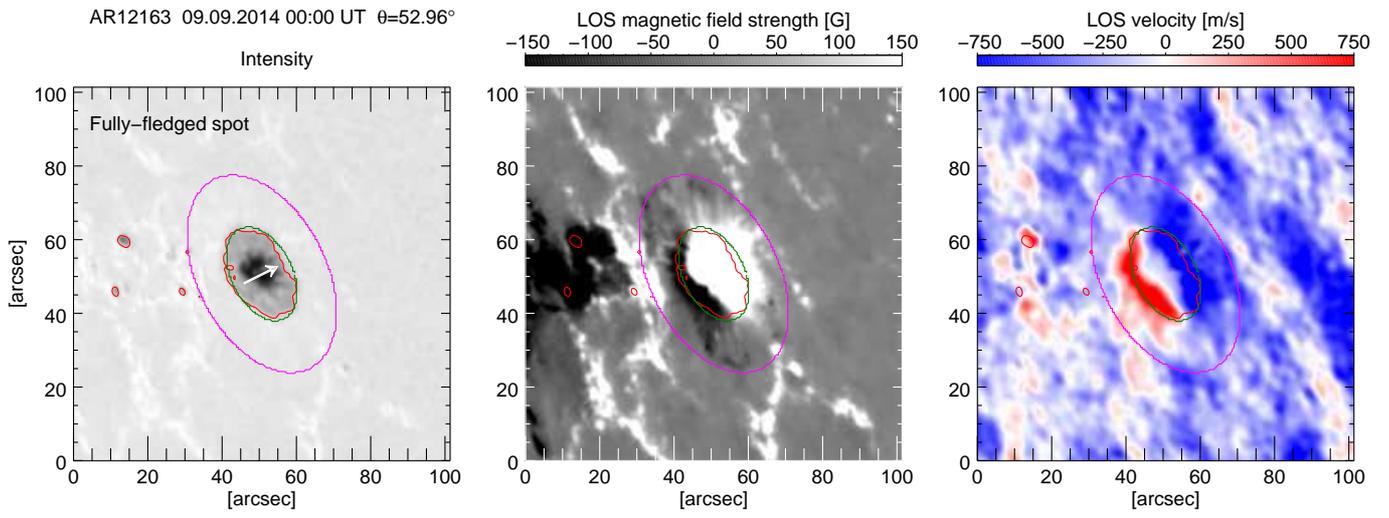}
\caption{Same as Fig.\,\ref{fig:ar1_fig}, except for AR12163.} 
\label{fig:ar6_fig}
\end{figure*}

\begin{figure*}[!h]
\centering
\includegraphics[width=1\textwidth]{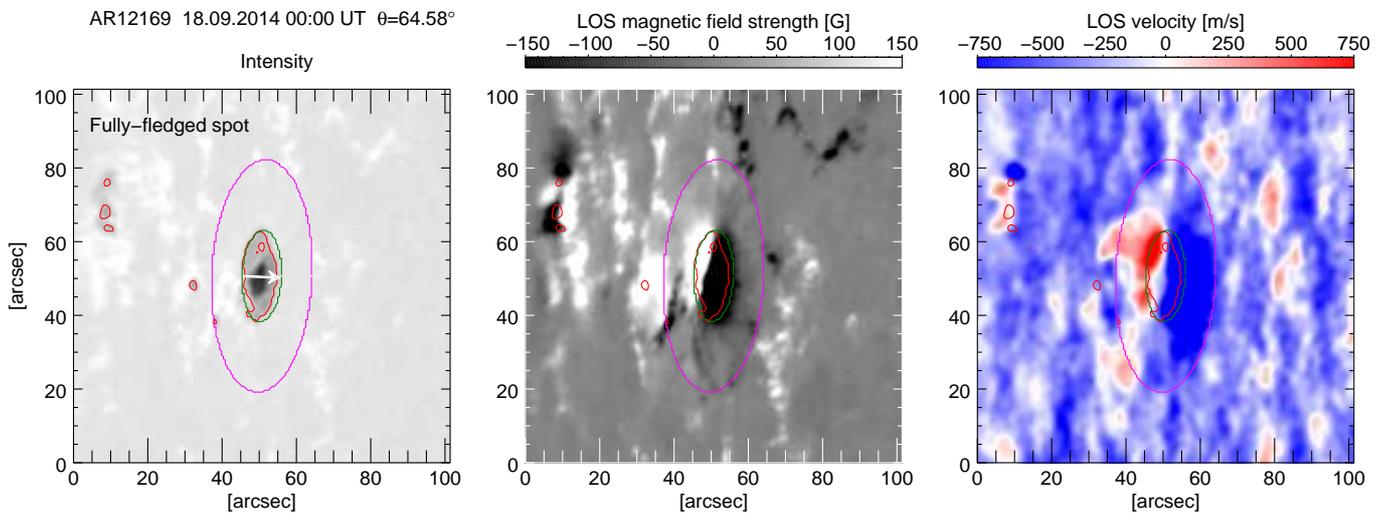}
\caption{Same as Fig.\,\ref{fig:ar1_fig}, except for AR12169.} 
\label{fig:ar7_fig}
\end{figure*}

\begin{figure*}[!h]
\centering
\includegraphics[width=1\textwidth]{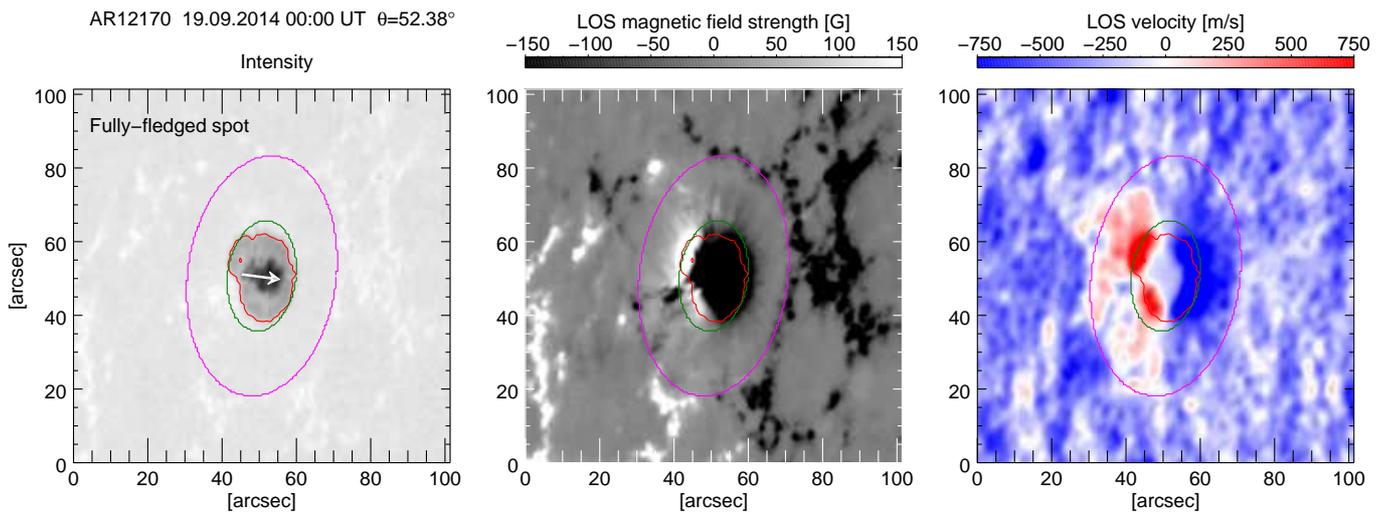}
\caption{Same as Fig.\,\ref{fig:ar1_fig}, except for AR12170.} 
\label{fig:ar8_fig}
\end{figure*}

In conclusion, sunspots embedded in a moat cell decay into naked spots embedded in a supergranule. The supergranule can remain after the naked spot has dissolved or becomes squeezed by surrounding supergranular cells, in the very same manner in which standard supergranular cells compete against each other.
%===================================================================================
 %                                ACKNOWLEDGEMENTS
%===================================================================================
\begin{acknowledgements}
 HS likes to thank J. L\"ohner-B\"ottcher for sharing his codes and for the helpful discussions. The authors thank R. Schlichenmaier for the careful reading of the manuscript and discussions.
HS has been funded by SOLARNET, an EU-FP7 integrated activity project and the DFG project RE 328 2/1-2.
 NBG acknowledges financial support by the Senatsausschuss of the Leibniz-Gemeinschaft, Ref.-No. SAW-2012-KIS-5 within the CASSDA project.
 SDO is a mission for NASA's Living With a Star (LWS) Program. This research has made use of NASA's Astrophysics Data System.
\end{acknowledgements}
\bibliographystyle{aa}
\bibliography{bibfile_mf.bib}
\begin{appendix}
\onecolumn\noindent\begin{minipage}[l]{0.49\textwidth}\section{Snapshots and animations of sunspot evolution}\label{app:spots}
The snapshots (see Fig.\,\ref{fig:ar1_fig} - \ref{fig:ar8_fig}) show the first time step of the analysed data set for each of the eight sunspots  in the intensity map (left), LOS magnetograms (middle), and LOS Doppler maps (right), as listed in Table\,\ref{tab:list_sunspots}. For each of the spots, an animation with the full temporal evolution of the active regions\end{minipage}\hspace{\fill}\noindent\begin{minipage}[r]{0.49\textwidth}is provided as online material. The spot boundary is outlined by a red contour at each time. The position of the maximum horizontal velocity and extension of the horizontal flow, that is, the cell size as determined with our analysis method, are marked by green and pink contours, respectively. The white arrow in the intensity map points in direction of disc centre and therein visualises the movement of the spot across the solar disc.\end{minipage}
%%%%%%%%%%%  FIGURE 4 %%%%%%%%%%%%%%%

%%%%%%%%%%%%%%%%%%%%%%%%%%%%%%%%%%%%%%%%%%%%%%%%%%%%%
\clearpage
\section{Figures of supergranules}\label{app:sg}
%%%%%%%%%%%  FIGURE 5 %%%%%%%%%%%%%%%
\begin{figure*}[!h]
\centering 
\includegraphics[width=1\textwidth]{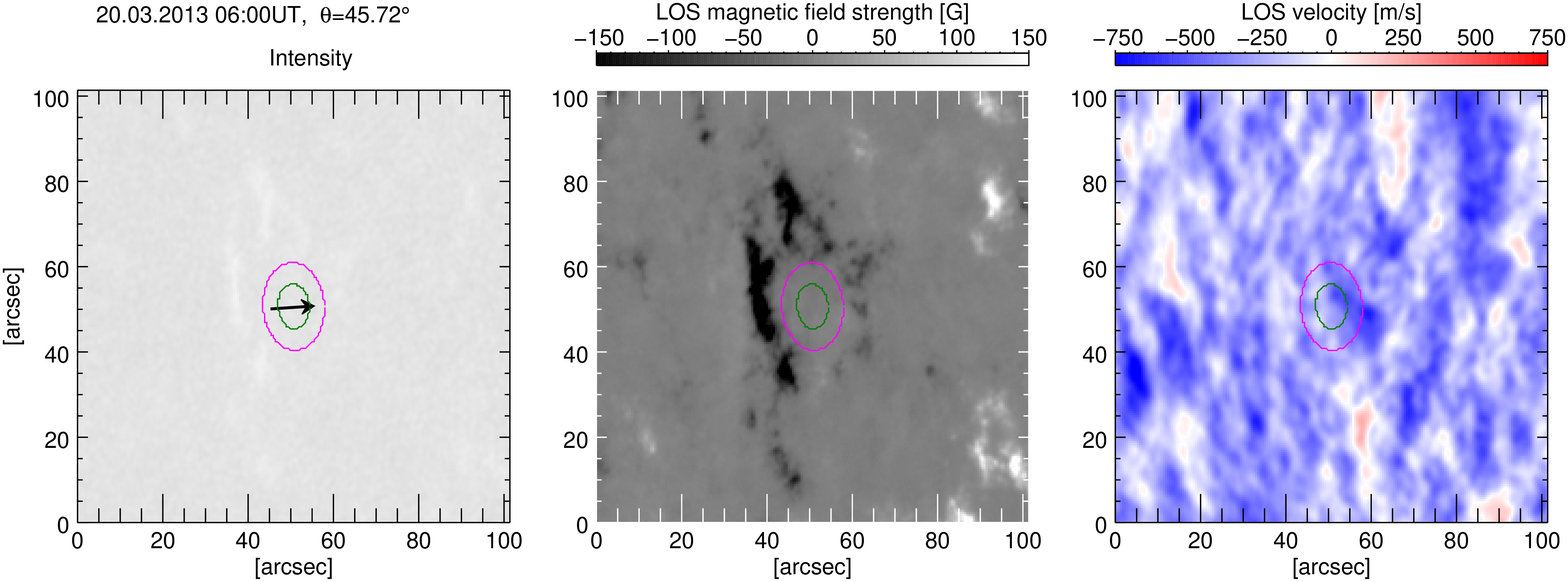}
\includegraphics[width=1\textwidth]{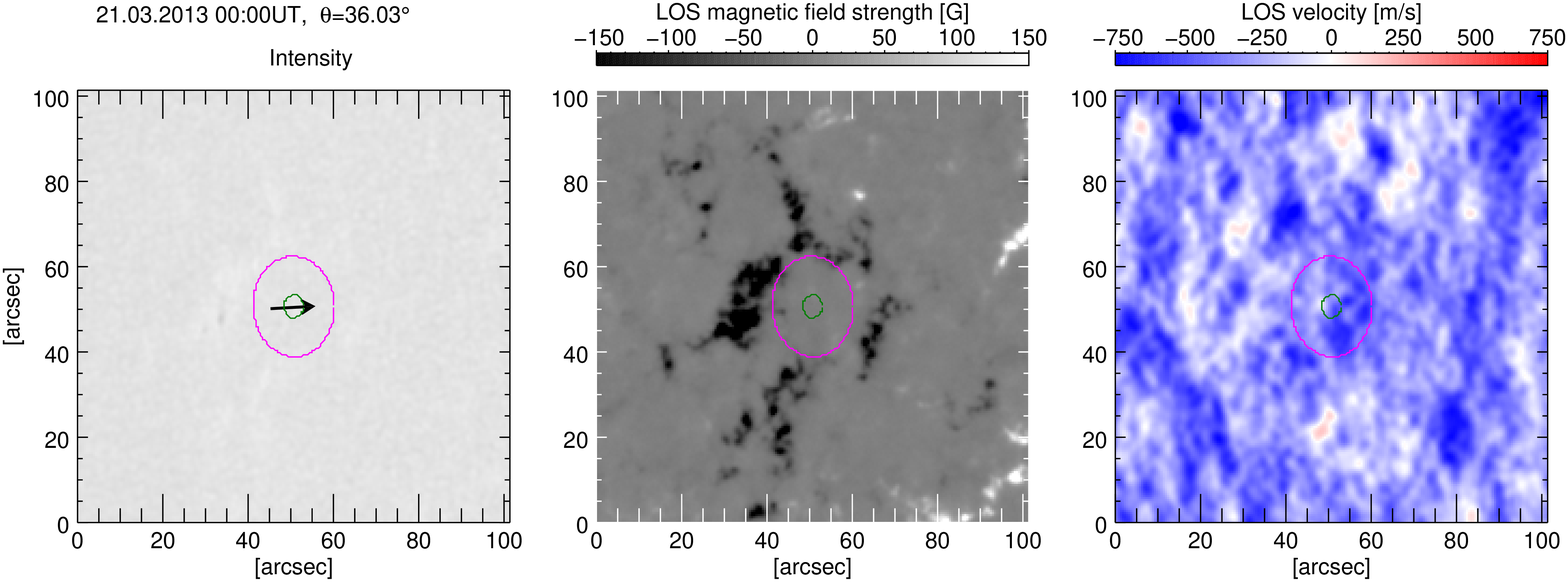}
\includegraphics[width=1\textwidth]{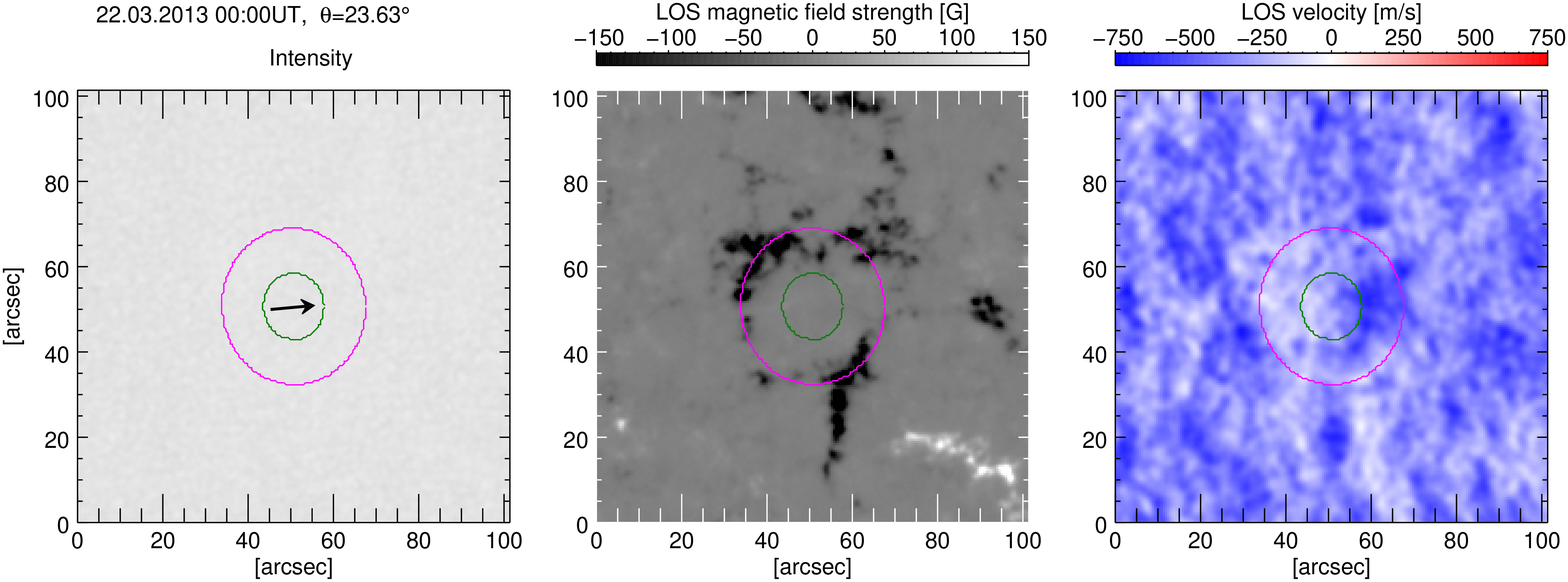}
\caption{Snapshots of three supergranules in intensity map{s} (left), LOS magnetograms (middle), and LOS Doppler maps (right). The position of the maximum horizontal velocity and extension of the horizontal flow, i.e., the cell size as determined with our analysis method, is marked by green and pink contours, respectively. The black arrow in the intensity map points in the direction of disc centre.} 
\label{fig:sg_fig}
\end{figure*}
%%%%%%%%%%%  FIGURE 5 %%%%%%%%%%%%%%%
\end{appendix}

\end{document}